\documentclass[twocolumn, twocolappendix]{aastex631}

\usepackage{amssymb}
\usepackage{amsmath}

\newcommand{\vect}[1]{\boldsymbol{#1}}

\shorttitle{Tides in the migration of hot Jupiters}
\shortauthors{Yu, Weinberg, \& Arras}
\begin{document}

\title{Tidal evolution and diffusive growth during  high-eccentricity planet migration:\\ 
revisiting the eccentricity distribution of hot Jupiters}

\correspondingauthor{Hang Yu}
\email{hangyu@caltech.edu}

\author[0000-0002-6011-6190]{Hang Yu}
\affiliation{TAPIR, Walter Burke Institute for Theoretical Physics, Mailcode 350-17 California Institute of Technology, Pasadena, CA 91125, USA}

\author[0000-0001-9194-2084]{Nevin N. Weinberg}
\affiliation{Department of Physics, University of Texas at Arlington, Arlington, TX 76019, USA}

\author[0000-0001-5611-1349]{Phil Arras}
\affiliation{Department of Astronomy, University of Virginia, P.O. Box 400325, Charlottesville, VA 22904, USA}




\begin{abstract}
High-eccentricity tidal migration is a potential formation channel for hot Jupiters. During this process, the planetary f-mode may experience a phase of diffusive growth, allowing its energy to quickly build up to large values. In \citet{Yu:21}, we demonstrated that nonlinear mode interactions between a parent f-mode and daughter f- and p-modes expand the parameter space over which the diffusive growth of the parent is triggered. We extend that study by incorporating  (1) the angular momentum transfer between the orbit and the mode, and consequently the evolution of the pericenter distance, (2) a phenomenological correction to the nonlinear frequency shift at high parent mode energies, and  
%so that the fractional shift does not exceed unity when the parent's energy builds up; 
(3) dissipation of the parent's energy due to both turbulent convective damping of the daughter modes and strongly nonlinear wave-breaking events. The new ingredients allow us to follow the coupled evolution of the mode and orbit over $\gtrsim 10^4$ years, covering the diffusive evolution from its onset to its termination.  
We find that the semi-major axis shrinks by a factor of nearly ten over $10^4$ years, corresponding to a tidal quality factor  $\mathcal{Q}\sim10$.
%about 4 orders of magnitude smaller compared to the empirical value determined from the Jupiter-Io system. 
%The f-mode's diffusive growth terminates while the ecccentricity is still high ($e\simeq 0.8-0.9$), and although a subsequent tidal process is required to fully circularize the planet, we show that the rapidity of diffusive tidal evolution can explain the paucity of super-eccentric ($e>0.9$) systems in the  \emph{Kepler} sample.
%\nevin{Instead of the previous ending: 
The f-mode's diffusive growth  terminates while the eccentricity is still high, at around $e=0.8-0.95$. Using these results,  we revisit the eccentricity distribution of proto-hot Jupiters. We estimate that less than 1 proto-HJ with eccentricity $>0.9$ should be expected in \emph{Kepler}'s data once the diffusive regime is accounted for, explaining the observed paucity of this population. 
%\hang{(HY: I think it is okay to not cite Wu in the abstract as long as we cite and credit her in the main text.)}
%} \phil{(cite Wu in abstract, since she already came to this conclusion?)}

\end{abstract}



\section{Introduction} 
\label{sec:intro}

In the standard scenario, tidal friction in a binary system arises when the motions induced by a time-changing tidal force are damped away irreversibly to heat within the bodies. The time variation may be due to either an eccentric orbit or an asynchronous or misaligned spin as compared to the orbital angular momentum. As encapsulated by Darwin's theory \citep{Darwin:1879}, friction causes a lag in the response of the body as compared to the driving force, and the forces from the resulting misaligned tidal bulge lead to secular exchange of energy and angular momentum between the two bodies and the orbit. Analytic formulas for the evolution of the orbit and spins are typically parametrized with a tidal ${\cal Q}$ or lag time $\tau$ (e.g., \citealt{Goldreich:66, Hut:81}), or attempts are made to  calculate the dissipation from first principles.

For the case of highly eccentric orbits, \citet{Mardling:95} suggested a qualitatively different route for tidal evolution as compared to Darwin's theory. In \citet{Mardling:95}, the pericenter distance is assumed to be so close that the tidal ``kicks" applied near pericenter can excite internal oscillation modes to significant energies. The energy transfer alters the orbit and changes its period. Importantly, these changes in orbital period may be so large that the phase of the oscillation mode at each pericenter passage is effectively random compared to the last, which will lead to a random walk in mode energy. On average, the mode amplitude (energy) grows as the square root (linearly) with the number of pericenter passages. This behavior is called the ``diffusive" or ``chaotic" tide.

If the system is conservative, the diffusive tide allows the mode energy to build up until the coupled orbital and mode degrees of freedom reach an approximate equipartition~\citep{Mardling:95, Vick:18}. If dissipation is involved and keeps removing energy from the mode, then on average there will be a net energy flowing continuously from the orbit to the mode. This may lead to a phase of rapid orbital evolution in which the orbital energy changes by many times its initial value and the semi-major axis decreases by factors of a few.

The diffusive tide may be an important process in the  high-eccentricity migration of proto-hot Jupiters (HJs). In this formation scenario for HJs, the planet is born beyond the snow line at $\gtrsim 1 \textrm{AU}$, its eccentricity is
driven to high values through interactions with another planet or star~(see, e.g., \citealt{Rasio:96,Wu:03, Fabrycky:07, Chatterjee:08, Wu:11, Teyssandier:13,  Petrovich:16, Munoz:16, Hamers:17}), and then
strong tidal dissipation in the planet damps the eccentricity and shrinks the orbital semi-major axis. If one simply applies \citeauthor{Darwin:1879}'s theory to this channel, it would predict a large number of proto-HJs with high orbital eccentricities~\citep{Socrates:12}. The
lack of observed systems in the high eccentricity state \citep{Dawson:15} thus leads to a tension between the theory and observation, seemingly disfavoring the high-eccentricity migration channel. 

Nevertheless, the tension could be naturally reconciled with the incorporation of the diffusive tide, as demonstrated in a number of studies~\citep{Ivanov:04, Ivanov:07, Wu:18, Vick:18, Vick:19}. In particular, \citet{Wu:18} showed that the $l=2$ f-mode of the planet can grow rapidly to very significant energies. If this energy can be further dissipated somehow via nonlinear processes, it can then lead to rapid orbital evolution on $\sim 10^4\textrm{ yr}$ timescales.  This would correspond to a tidal quality factor $\mathcal{Q}\sim 1-10$, which is about four to five orders of magnitude smaller than Jupiter's. \citet{Wu:18} showed that this could explain the paucity of highly eccentric systems found by \emph{Kepler} because the circularization is so fast during the high-eccentricity
phase that catching one in this state is observationally unlikely. 

Previous studies considered the diffusive tide in the linear approximation.  In this approach, the source of randomness of the f-mode phase is tidal back-reaction on the orbit.  In our previous work \citep{Yu:21}, we extended the analysis by including weakly nonlinear mode interactions between the parent f-mode (the mode that couples directly to the tide) and daughter f- and p-modes. We showed that the mode interactions lead to an energy-dependent shift of the parent f-mode's oscillating frequency. As a result, they provide  another channel for randomizing the f-mode's phase, in addition to the tidal back-reaction considered by previous studies.
The phase shift produced by the two channels are formally the same order in terms of the energy transfer at each pericenter passage and therefore have comparable significance in triggering the diffusive tide. However, once typical parameters of a proto-HJ are plugged in, we showed in \citet{Yu:21} that the phase shift due to nonlinear mode interactions is $\approx 5$ times greater than that due to tidal back-reaction on the orbit. Therefore, it is nonlinear interactions that dominate the triggering and subsequent maintenance  of the diffusive tidal evolution in migrating proto-HJs. 

In \citet{Yu:21}, we only considered the parameter space that can trigger the diffusive tide. We extend the study here by now considering the long-term evolution of the system over $\sim 10^4\,{\rm yr}$. In Section~\ref{sec:formalism}, we describe the formalism and extensions to \citet{Yu:21} that allow us to solve for the long term evolution.  In Section~\ref{sec:results} we present the results of evolving the coupled mode and orbit equations over $\sim 10^4\textrm{ yr}$ including nonlinear mode interactions.  But first, in Section~\ref{sec:number_along_track},  we show the impact of the diffusive tide on the expected number density of migrating proto-HJs and compare the results with the observed population.

\section{Eccentricity distribution of migrating proto-hot Jupiters}
\label{sec:number_along_track}

\begin{figure}
   \centering
   \includegraphics[width=0.45\textwidth]{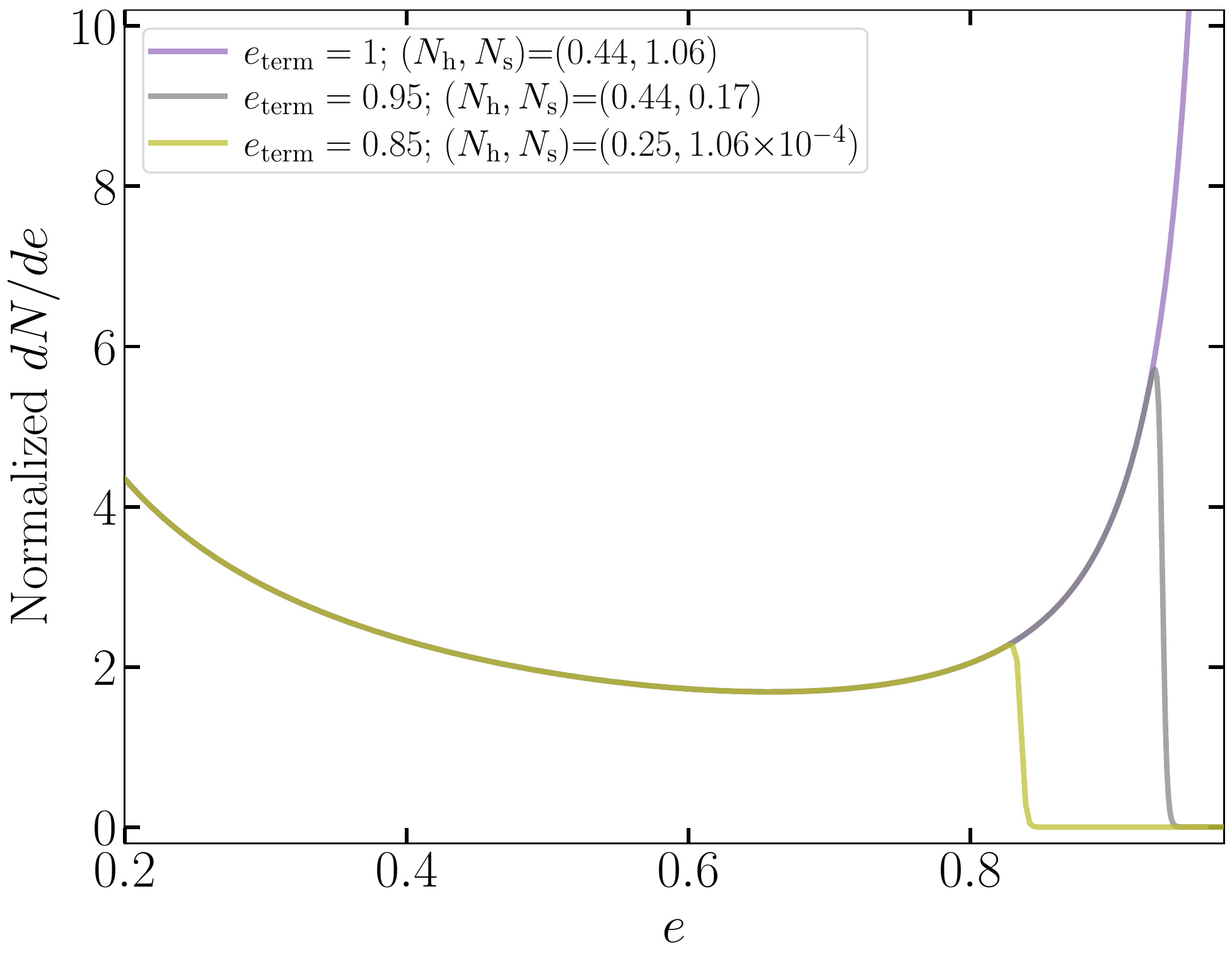} % requires the graphicx package
\caption{Number density as a function of the orbital eccentricity of migrating proto-hot Jupiters. We show in the legend the number of highly-eccentric (denoted by $N_{\rm h}$; corresponding to $0.7<e<0.9$) or super-eccentric (denoted by $N_{\rm s}$, $0.9<e<e_{\rm max}$) systems relative to mildly-eccentric systems (denoted by $N_{\rm m}$; corresponding to $0.2<e<0.6$). Line colors  indicate different assumptions about the eccentricity $e_{\rm term}$ at which diffusive growth terminates with purple, gray, and olive lines corresponding to $e_{\rm term}=(1, 0.95, 0.85)$, respectively. 
}
\label{fig:num_vs_e}
\end{figure}

A key result of our study is the influence of the diffusive tide on the eccentricity distribution of migrating proto-HJs. 
%\hang{(HY: let us use HJs in the paper. The abbreviation ``HJ'' needs to be defined in the intro.)}  
To evaluate this, we follow the prescription outlined in \citet{Socrates:12} but generalize the treatment to allow two phases of orbital evolution. During the initial phase, when the diffusive tide operates, the orbital energy is efficiently removed and the orbital evolution is fast. Later, when the diffusive growth terminates (whose condition we will explore in later sections), the orbital evolution slows considerably.. %which we approximate under the same phenomenological model assumed by \citet{Socrates:12}.

Quantitatively, we assume there is a constant current of migrating planets, and therefore 
\begin{equation}
     \frac{dN(e; L)}{de}\frac{d e}{dt} = {\rm const.},
\end{equation}
where $N(e; L)$ is the total number of planets with orbital eccentricity less than $e$ and orbital angular momentum (AM) $\simeq L$. 
%\hang{(HY: I would slightly prefer to keep the $L$ in the subscript as it is a parameter entering the calculation. Systems have different $L$ will have slightly different $dN_L/de$.)} \nevin{ok with me} 
We assume the planet migrates along a track of constant $L$, which is a good approximation since the transfer of orbital AM into the planetary f-mode by the diffusive tide changes $L$ by less than $1\%$ (see Sections~\ref{sec:orbital_AM} and \ref{sec:results_1p1_Rj}). 
At any time, the number density of migrating planets with angular momentum $L$ at eccentricity $e$ is therefore given by $dN/de \propto 1/|de/dt|$. To calculate $de/dt$, we follow the phenomenological approach of \citet{Hut:81} and assume the planet is pseudo-synchronized, which implies\footnote{Two caveats exist when adopting the model by \citet{Hut:81}. The first is that after the diffusive tide terminates, we still approximate the tidal lag time as a constant. This is not necessarily a good assumption as the tidal lag time is not a fundamental property of the planet and it can depend sensitively on the driving frequency (see, e.g., \citealt{Ogilvie:04, Weinberg:12}). In fact, what drives the evolution from $e\sim 0.8$ to nearly zero eccentricity remains unclear. The second is that we set the rotation rate according to the pseudo-synchronization value. This may be inaccurate during the diffusive evolution because of the AM transferred to the f-mode. While it is small compared to $L$, it can nonetheless be significant for the planet. Evaluating the importance of both caveats is left for future study.}
\begin{equation}
    \frac{de}{dt} = -\frac{1}{2 T(e)} e(1-e^2)^{3/2} g(e),
\end{equation}
where 
\begin{equation}
    g(e) = \frac{7 + \frac{45}{2}e^2 
                     + 56 e^4 + \frac{685}{32}e^6
                     + \frac{255}{64} e^8 + \frac{25}{256}e^{10}}
                     {3(1+3e^2+\frac{3}{8}e^4)},
\end{equation}
and 
\begin{eqnarray}
    T(e) &= \frac{M a^8\left(1-e^2\right)^8}{9k_2GM_\ast(M+M_\ast)R^5 \tau(e)}\nonumber \\
         &= \frac{\left(M+M_\ast \right)^7 L^{16}}{9 k_2 G^9 M^{15}M_\ast^{17}R^5\tau(e)}.
    \label{eq:T_tidal_scale}
\end{eqnarray}
Here $M_\ast$, $a$, $M$, $R$, and $k_2$ are, respectively, the mass of the host star, the orbital semi-major axis, and the mass, radius, and Love number of the planet. We approximate the orbital evolution as a two-phase process by letting the tidal time lag be a step function, 
\begin{equation}
    \tau(e) = 
    \begin{cases}
        0.1\,{\rm s} & \text{for } e \leq e_{\rm term}, \\
        10^3\,{\rm s} & \text{for } e> e_{\rm term},
    \end{cases}
    \label{eq:tau_tidal_lag}
\end{equation}
where the diffusive evolution terminates at $e_{\rm term}$. Since large $\tau$ implies large $|de/dt|$, the tidal dissipation is four orders of magnitude more efficient when  $e>e_{\rm term}$ and the  diffusive tide operates. By changing the value of $e_{\rm term}$, we can further examine how the terminal eccentricity affects the eccentricity distribution of migrating planets. 
The values of $e_{\rm term}$ and $\tau(e)$ when $e> e_{\rm term}$ will be justified in later sections. 
For $e<e_{\rm term}$, we set $\tau=0.1\,{\rm s}$, which is motivated by the empirically measured dissipation in the Jupiter-Io system (see, e.g., \citealt{Vick:19}). 

In Figure~\ref{fig:num_vs_e} we examine three cases. In the purple curve, we set $e_{\rm term}=1$, which corresponds to the case without the diffusive tide; it reduces to the scenario considered in \citet{Socrates:12}. The gray and olive curves include the diffusive tide when $e >  e_{\rm term}$, with the former having  $e_{\rm term}=0.95$ and the later having $e_{\rm term}=0.85$. As we will see later in Section~\ref{sec:results}, they correspond to the typical termination point of the diffusive tidal evolution under the linear theory and including nonlinear mode interactions, respectively.

% One with $e_{\rm term}=0.95$ (gray curve), corresponding to the value where the diffusive tidal evolution terminates under the linear theory, and one with $e_{\rm term}=0.85$ (olive curve) for the typical terminal point once nonlinear interactions are included (see Section~\ref{sec:results}). Lastly, we also consider a case with $e_{\rm term}=1$, which reduces to the case considered in \citet{Socrates:12} without diffusive tide at all. 
%All trajectories have a final orbital period of 3 days. A trajectory with final period of 5 days will be also similar to the ones shown here. 

Given the number density $dN/de$, we can further compute the number of systems in a given range of eccentricities $(e_1, e_2)$ by computing $\int_{e_1}^{e_2} \left[dN(e)/de\right]de$. Specifically, we define systems with $0.2<e<0.6$ as mildly-eccentric and denote the number of such systems as $N_{\rm m}$. As suggested in \citet{Socrates:12} (see also \citealt{Dawson:15}), we can use this population to calibrate various constants entering the calculation and use it to predict the number of planets at different eccentricities. We are particularly interested in $N_{\rm h}$, the number of highly-eccentric proto-HJs (defined as $0.7<e<0.9$), as well as $N_{\rm s}$ the number of super-eccentric systems ($0.9<e<e_{\rm max}$). %\footnote{To calculate $N_{\rm s}$ for the super-eccentric population, we additionally put an upper cut of $e_2=0.98$, corresponding to an initial $a_{\rm orb, 0}=1\,{\rm AU}$ for our migration track. } 
Here $e_{\rm max}$ is the maximum eccentricity to which a survey is sensitive.  We consider  in this study the number of proto-HJs in \emph{Kepler}'s data \citep{Burke:14} and determine $e_{\rm max}$ following \citet{Dawson:15},
\begin{equation}
    e_{\rm max} = \left\{1 - \left[\frac{(n_{\rm trans}-1)P_{\rm orb}^{\rm (f)}}{t_{\rm survey}}\right]^{2/3}\right\}^{1/2},
\end{equation}
where $n_{\rm trans}$ is the minimum number of transits required for detection, $t_{\rm survey}$ is the duration of the mission, and $P_{\rm orb}^{\rm (f)}$ is the final (when $e=0$) orbital period. 
%Physically, it is determined by finding $P_{\rm orb}=t_{\rm survey}/(n_{\rm trans}-1)$ when $e=e_{\rm max}$ so that during the observation $n_{\rm trans}$ transits can be observed. 
The expression for $e_{\rm max}$ follows by noting that for constant AM, $P_{\rm orb}(1-e^2)^{3/2}=P_{\rm orb}^{\rm (f)}$, and then using the fact that the maximum orbital period for which a survey can detect at least $n_{\rm trans}$ transits in a time $t_{\rm survey}$ is $P_{\rm orb} = t_{\rm survey}/(n_{\rm trans}-1)$.
As in \citet{Dawson:15}, we use $n_{\rm trans}=3$ and $t_{\rm survey}=4\,{\rm yr}$ in our calculation. Note here we consider only the \emph{intrinsic} population and do not include selection effects other than the cut in $e_{\rm max}$. In the plot, we set the normalization such that $N_{\rm m}=1$ and the value of $(N_{\rm h}, N_{\rm s})$ are shown in the legend for each curve. The curves are generated for $P_{\rm orb}^{\rm (f)}=3\,{\rm day}$. Changing the final orbital period only affects the value of $N_{\rm s}$ for the case without diffusive tide (i.e., the purple curve). For $P_{\rm orb}^{\rm (f)}=5 (10)\,{\rm day}$, we have $N_{\rm s} = 0.82 (0.55)$ for the purple curve. Approximately, we have $N_{\rm s}\propto \left[P_{\rm orb}^{\rm (f)}\right]^{-1/2}$. Other values depend weakly on $P_{\rm orb}^{\rm (f)}$. 

If we ignore the diffusive tide (purple curve), we see that there should be a significant number of super-eccentric planets, as predicted by \citet{Socrates:12}.
Indeed, \citet{Dawson:15} estimated that, after correcting for selection effects, there should be about 1.5 (3) super-eccentric proto-HJs with $3\,{\rm day}<P_{\rm orb}^{\rm (f)}<5\,{\rm day}$ ($5,{\rm day}<P_{\rm orb}^{\rm (f)}<10\,{\rm day}$) in \emph{Kepler}'s data.  However, \citet{Dawson:15} found only $N_{\rm s}=0_{-0}^{+1}$ in the data. %\phil{(this $0\pm 1$ is kind of awkward. maybe just explain data in words. why is there an error bar?)} 
(figure 2 and section 3.2 in \citealt{Dawson:15}). On the other hand, including the diffusive tide can reduce this number by a significant factor. Comparing the gray curve to the purple one, we find that the number of super-eccentric proto-HJs is reduced by a factor of $6, 5, 3$ for $P_{\rm orb}^{\rm (f)}=3, 5, 10\,{\rm day}$ if $e_{\rm term}=0.95$. If the diffusive tide could be maintained to an eccentricity $e_{\rm term}=0.85$ as shown in the olive curve, then it should be reduced by a factor of $\sim 10^4$. These estimates are consistent with those of  \citet{Wu:18}, who considered the linear diffusive tide. The influence of the diffusive tide can therefore naturally explain the paucity of super-eccentric proto-HJs. %\phil{(What is the maximum $P_{\rm orb}^{(f)}$ for which diffusive growth occurs? Could it extend out to 10 days?)}

In order to further constrain the details of tidal migration and distinguish between, e.g., the gray and olive curves, this would require detecting a population of highly-eccentric proto-HJs with $0.7<e<0.9$. Assuming the gray curve (with $e_{\rm term}=0.95$) corresponds to the true population, we estimate that about 10 highly-eccentric systems would be needed in order for the statistical uncertainty $\sim \sqrt{10}$ to be smaller than the difference between the gray and the olive curves. Current observations do not constrain this band well \citep{Dawson:15} and future surveys are needed. Alternatively, a detection of a proto-HJ with $e\simeq 0.9$ and a pericenter distance sufficiently small to trigger diffusion during its evolution~\citep{Wu:18, Yu:21} could also distinguish the gray and the olive curves.\footnote{We note that HD 80606b  has an eccentricity of $e=0.983$ \citep{Wu:03}. However, its pericenter distance is too large to trigger diffusive growth given its radius of $\simeq 0.9 R_J$. See also the discussion in \citet{Wu:18}. This suggests that the number density of proto-HJs is in fact more complicated than the framework proposed in \citet{Socrates:12} which we adopt here. Instead, the properties of the planet itself, such as its radius, could also play crucial roles (see discussions in Appendix~\ref{appx:P_f_vs_R}). Further investigations are thus required.  }

\section{Formalism for the dynamical tide}
\label{sec:formalism}
In the previous section we, illustrated the key observational consequence of the diffusive tide, i.e., explaining the observed paucity of super-eccentric proto-HJs.  In this section, we first review the key ingredients needed to construct an iterative mapping to evolve the tidal amplitude equation, following the framework outlined in \citet{Yu:21} (see also \citealt{Ivanov:04, Wu:18, Vick:18, Vick:19}). We then describe additional physics not accounted for in our original analysis, including orbital AM transfer,  higher order corrections to the frequency shift, and strongly nonlinear energy dissipation (Sections~\ref{sec:orbital_AM}, \ref{sec:NL_freq_shift}, and \ref{sec:energy_dissipation}, respectively). 

Suppose the host star (of mass $M_\ast$) raises a tide in the planet (of mass $M$) with a Lagrangian displacement $\vect{\xi}(\vect{x}, t)$. We can expand $\vect{\xi}(\vect{x}, t)$ as 
\citep{Schenk:02, Weinberg:12}
\begin{equation}
\left\{
\begin{array}{l}
\vect{\xi}(\vect{x},t) = \sum_a q_a(t) \vect{\xi}_a(\vect{x},t) \\
\dot{\vect{\xi}}(\vect{x},t) = \sum_a (-i \omega_a) q_a(t) \vect{\xi}_a(\vect{x},t), \\
\end{array}
\right.
\end{equation}
where $q_a(t)$ is the amplitude of an eigenmode and $\omega_a$ is its eigenfrequency. The sums run over both radial and angular quantum numbers as well as modes with both positive and negative frequencies. 
We normalize each mode such that 
\begin{equation}
    2\omega_a^2\int d^3 x \rho \vect{\xi}^\ast_{a'}\cdot\vect{\xi}_a = E_0\delta_{aa'},
\end{equation}
where $E_0=GM^2/R$.  
For a parent mode that directly couples to the tide, its evolution in the co-rotating frame is described by the amplitude equation~\citep{Yu:21}
\begin{equation}
\dot{q}_a + \left[i (\omega_a + \delta \omega_a)  + (\gamma_a+ \delta \gamma_a)\right] q_a 
	= i \omega_a U_a,  \label{eq:ode_mode_amp}
\end{equation}
where the linear tidal forcing is given by
\begin{equation}
   U_a(t) =W_{lm} Q_{a} \left( \frac{M_\ast}{M} \right)  \left[\frac{R}{D(t)}\right]^{l+1} \exp\left\{-i m \left[\Phi(t) - \Omega_{\rm s}t\right]\right\},
\label{eq:U_a} 
\end{equation}
where  $D$ is the orbital separation, $\Phi$ is the orbital phase, $\Omega_{\rm s}$ is the spin of the planet, the linear tidal overlap $Q_a=(MR^l)^{-1}\int d^3r\rho \vect{\xi}^\ast \cdot \nabla (r^l Y_{\rm lm}) $, and at leading order ($l=2$), the nonvanishing $W_{lm}$ coefficients are $W_{2\pm2}=\sqrt{3\pi/10}$ and $W_{20}=-\sqrt{\pi/5}$. The quantities $\omega_a$ and $\gamma_a$ are the linear eigenfrequency and damping rate of the mode, and $\delta \omega_a$ and $\delta \gamma_a$ are the nonlinear corrections caused by the parent mode (an $l_a=2$ f-mode) interacting with high-order daughter f- and p-modes~\citep{Yu:21}. For small mode energy $\tilde{E}_a \equiv  |q_a|^2 (<|\omega_a/\Omega|<1)$, the nonlinear corrections can be expressed as~\citep{Yu:21}
\begin{eqnarray}
    &\delta \omega_a = \Omega \tilde{E}_a, \label{eq:NL_freq_shift_leading_order}\\ 
    &\delta \gamma_a = \Gamma \tilde{E}_a,
    \label{eq:NL_damp_shift}
\end{eqnarray}
where $\Omega$ and $\Gamma\propto \omega_a$ are constants determined by the internal structure of the planet. The typical values of $\Omega$ and $\Gamma$ will be presented later in Table~\ref{tab:models}. %\nevin{Maybe remind reader here what the typical values of $\Omega$ and $\Gamma$ are?} 
Here, as in \citet{Yu:21}, we use the tilde symbol to indicate energy normalized by $E_0$. For instance, $\tilde{E}_{{\rm orb}, k} = E_{{\rm orb}, k}/E_0$.

Upon a Doppler frequency shift (and ignoring other rotational corrections for simplicity), we can also transfer the co-rotating frame mode amplitude equation (Equation~(\ref{eq:ode_mode_amp})) to the inertial frame,
\begin{equation}
\dot{q}_a' + \left[i (\omega_a' + \delta \omega_a)  + (\gamma_a+ \delta \gamma_a)\right] q_a' 
	= i \omega_a U_a',  \label{eq:ode_mode_amp_inertial}
\end{equation}
where $q_a'=q_a\exp\left(-i m_a \Omega_{\rm s}t\right)$, $\omega_a'=\omega_a + m_a\Omega_{\rm s}$, and $U_a'=U_a\exp\left(-i m_a \Omega_{\rm s}t\right)$.

When the orbit is highly eccentric and the tidal interaction is only significant near each pericenter passage, Equation~(\ref{eq:ode_mode_amp_inertial}) can be evolved together with the orbit via a set of iterative mapping equations (assuming weak damping; the more general form will be presented in Section~\ref{sec:NL_freq_shift})
\begin{eqnarray}
    &&q_{a, k}'^{\rm (0)} {=} q_{a, k-1}'^{\rm (1)} + \Delta q_{a, k} \label{eq:iter_map_interation}\\
    &&q_{a, k}'^{\rm (1)} {=} q_{a, k}'^{\rm (0)} e^{-\left[i(\omega_a'+\Omega \tilde{E}_{a, k}^{(0)} )+(\gamma_a + \Gamma \tilde{E}_{a, k}^{(0)})\right] P_{{\rm orb}, k}},
    \label{eq:iter_map_mode_prop}\\
    &&\tilde{E}_{{\rm orb}, k} - \tilde{E}_{{\rm orb}, k-1} {=} {-} (\tilde{E}_{a, k}^{(0)} - \tilde{E}_{a, k-1}^{(1)}) {=} {-} \Delta \tilde{E}_{a, k}, 
    \label{eq:iter_map_E_orb}\\
    &&\frac{P_{{\rm orb}, k}}{P_{{\rm orb}, k-1}} {=} \left(\frac{\tilde{E}_{{\rm orb}, k}}{\tilde{E}_{{\rm orb}, k-1}}\right)^{-3/2},
    \label{eq:iter_map_P_orb}
\end{eqnarray}
where the superscript $(0)$ and $(1)$ indicate that the amplitudes are, respectively, evaluated right after and right before a pericenter passage.  We have also defined $\tilde{E}_{a,k}^{(0)}=|q_{a, k}'^{\rm (0)}|^2$ and similarly for $\tilde{E}_{a, k-1}^{(1)}$.  The tidal interaction at each pericenter is characterized by a kick in the mode amplitude, $\Delta q_{a,k}$, given by 
\begin{eqnarray}
    \Delta q_{a,k} &=& \int i\omega_a U'_a(t) e^{(i\omega'_a+\gamma_a)t} dt, \nonumber \\
    &=&i 2\pi Q_a K_{lm}(\omega_a', \Omega_{\rm peri})  \nonumber \\
    &\times& \left(\frac{\omega_a\Omega_{\rm peri}^2}{\omega_0^3}\right) \left(\frac{M_\ast}{M+M_\ast}\right)\left(\frac{R}{D_{\rm peri}}\right)^{l-2},
    \label{eq:one_kick_amp}
\end{eqnarray}
where $D_{\rm peri}\equiv a(1-e)$ is the pericenter distance, $\Omega_{\rm peri}^2 \equiv G(M+M_\ast)/D_{\rm peri}^3$, and $\omega_0^2=GM/R^3$. The quantity $K_{lm}$, which corresponds to the temporal overlap between the orbit and the mode, is given by~\citep{Press:77}
\begin{eqnarray}
    &&K_{lm}(\omega, \Omega_{\rm peri}) = \frac{\omega_0 W_{lm}}{2\pi}\int  \left[\frac{D_{\rm peri}}{D(t)}\right]^{l+1} e^{i\left[\omega t - m\Phi(t)\right]} dt. \nonumber \\
    \label{eq:K_lm_def}
\end{eqnarray}
Assuming a nearly parabolic orbit and ignoring tidal back-reaction, one can further write $K_{22}$ (corresponding to the prograde $l=m=2$ harmonic which dominates the tidal interaction) as~\citep{Lai:97}
\begin{eqnarray}
     K_{22}(\omega, \Omega_{\rm peri})&\simeq&\frac{2z^{3/2}e^{-2z/3}}{\sqrt{15}}\left(\frac{\omega_0}{\Omega_{\rm peri}}\right)\left(1-\frac{\sqrt{\pi}}{4\sqrt{z}}\right), \nonumber \\
    &\simeq& 1.1\times10^{-2} \left(\frac{z}{11}\right)^{-5.7}
    \left(\frac{\omega_0}{\Omega_{\rm peri}}\right),
    \label{eq:K_lm_para}
\end{eqnarray}
where $z\equiv\sqrt{2}\omega/\Omega_{\rm peri}$. Note that in the second equality we have expanded the expression around $z=11$, which is a typical value for systems we consider in this study.  

%A quantity that is particularly interesting is the one-kick energy, 
For future convenience, we define the one-kick energy as
\begin{equation}
    \Delta \tilde{E}_1 \equiv |\Delta q_1|^2,
\end{equation}
which is the energy gained by the mode after the first pericenter passage (assuming the mode starts from zero energy). Note, however, that in the subsequent passages, the mode energy gain $\Delta \tilde{E}_{a, k}\neq |\Delta q_k|^2$. Instead, if the mode has accumulated an energy $\tilde{E}_{a, k}^{(0)}\gg \Delta \tilde{E}_1$, then the characteristic energy exchange is given by (see, e.g., \citealt{Wu:18})
\begin{equation}
    \Delta \tilde{E}_{a, k}\sim |q_{a, k}'^{\rm (0)}  \Delta q_{a,k}|\sim \sqrt{\tilde{E}_{a, k}^{(0)} \Delta \tilde{E}_1}.
    \label{eq:Delta_E_k}
\end{equation}
Here we approximated $\Delta q_k\simeq\Delta q_1={\rm constant}$. This is because the change in $D_{\rm peri}$ is small and it can therefore  be approximated as constant throughout the evolution (see, however, Section~\ref{sec:orbital_AM}).  Equations~(\ref{eq:one_kick_amp}) and (\ref{eq:K_lm_para}) then imply that the kick amplitude is also approximately constant.

Equations~(\ref{eq:iter_map_interation})-(\ref{eq:iter_map_P_orb}) provide a complete set of equations needed to evolve the system including leading-order nonlinear interactions. However, one should keep in mind that they have some potentially important limitations. Although $D_{\rm peri}$ is nearly constant for much of the planet's orbital evolution, the kick $\Delta q_k$ is especially sensitive to $D_{\rm peri}$ (Equation~(\ref{eq:K_lm_para})). Furthermore, the leading-order nonlinear correction to the frequency, Equation~(\ref{eq:NL_freq_shift_leading_order}), is only accurate for small parent energies. As the mode grows diffusively, it can reach $\tilde{E}_a\gtrsim |\omega_a/\Omega|$, and the frequency correction $\delta \omega_a \gtrsim \omega_a$, which is likely unphysical. As $\tilde{E}_a$ grows even further to $\tilde{E}_a\gtrsim 0.1$, the parent mode's displacement at the surface may reach a value $\simeq R$, 
and it may break due to strongly nonlinear processes~\citep{Wu:18}. 

In \citet{Yu:21}, we restricted the analysis to the initial phase of the evolution (the first few hundred years), when the issues noted above are not yet critical. Since we are now interested in following the system over a longer timescale ($\gtrsim 10^3\textrm{ years}$), we must account for them.  We describe our procedure for doing so in Sections~\ref{sec:orbital_AM}-\ref{sec:energy_dissipation}.

\subsection{Orbital AM transfer}
\label{sec:orbital_AM}
In~\citet{Yu:21} and other studies (e.g., \citealt{Vick:18}), the pericenter distance $D_{\rm peri}$ is kept constant during the evolution. This approximation is reasonable at early times because at each pericenter passage, the fractional AM transfer is much smaller than the fractional energy transfer (Appendix~\ref{appx:AM_transfer}) and during the highly eccentric stage, $\Delta D_{\rm peri}/D_{\rm peri}\simeq 2\left(\Delta L/L\right)$ since $L\simeq \mu\sqrt{2G(M+M_\ast)D_{\rm peri}}$, where $\mu=MM_\ast/(M+M_\ast)$ is the reduced mass. However, the strength of the tidal kick is very sensitive to the pericenter distance (Equation~(\ref{eq:K_lm_para}); see also \citealt{Wu:18}). We thus relax here the approximation that the pericenter distance is constant, and instead incorporate its evolution into the iterative mapping. 

Together with the energy transfer from the orbit to the planet $\Delta E_{a, k}=\left(\tilde{E}_{a, k}^{(0)} - \tilde{E}_{a, k-1}^{(1)}\right)E_0$, we also have an associated AM transfer $\Delta J_{a, k}$ given by (see  Appendix~\ref{appx:AM_transfer})
\begin{equation}
    \Delta J_{a, k}= L_{k} - L_{k-1} = \frac{m_a}{\omega_a}\Delta E_a. 
    \label{eq:dJ_eq_dL}
\end{equation}
We use $J_a$ to denote the AM of mode $a$ in order to distinguish it from the orbital AM $L$.
From the new orbital AM (together with $E_{{\rm orb},k}$ or, equivalently, $a_{k}$), we can then calculate the new orbital eccentricity as
\begin{equation}
    e_{k} = \left[1-\frac{L_k^2}{\mu^2G(M+M_\ast)a_k}\right]^{1/2},
    \label{eq:e_orb_from_L_orb}
\end{equation}
and the new pericenter distance as 
\begin{equation}
    D_{{\rm peri}, k} = \frac{a_{k}}{1-e_{k}}=\frac{L_{k}^2}{\mu^2G(M+M_\ast)(1+e_{k})}.
\end{equation}
We then update the kick amplitude in the next [the $(k+1)$'th] pericenter passage with the newly computed $D_{{\rm peri}, k}$.\footnote{In fact, the new pericenter distance should already affect the kick-amplitude at the $k$'th passage and thus $\Delta E_{a,k}$. However, the fractional change in $D_{\rm peri}$ is only $\mathcal{O}(10^{-6})$ after a single pericenter passage; it is only the cumulative change in $D_{\rm peri}$ over $\mathcal{O}(10^4)$ orbital cycles that plays a mildly important role. Therefore, shifting $D_{\rm peri}$ by a single cycle in our mapping procedure does not affect the qualitative results.}
%\footnote{In fact, the new pericenter distance should already affect the kick-amplitude at the $k$'th passage and thus $\Delta E_{a,k}$. One may account for this by re-evaluate $\Delta E_{a,k}$ and other quantities using the new $D_{{\rm peri},k}$ and iterative this process until convergence. Nonetheless, we find only the cumulative change in $D_{\rm peri}$ over $\mathcal{O}(10^3)$ cycles plays a mild role in the evolution. Therefore, shifting $D_{\rm peri}$ by a single cycle does not affect the qualitative results. } 

Note that we have effectively treated the planetary f-mode as a reservoir of the AM here. For simplicity, we do not consider in this study how the AM in the f-mode might be further deposited into the background planet as the mode energy is dissipated (either by strongly or weakly nonlinear processes). As a result, we do not change the spin of the background planet during the evolution. Furthermore, we ignore any AM transfer due to the equilibrium tide as the linear damping rate $\gamma_a$ is extremely small and hence the torque due to the equilibrium tide is very weak. As a caveat, we note that the spin evolution of the planet could nonetheless have significant impact on the evolution of the orbit because the kick at each pericenter passage depends sensitively on the mode's inertial-frame frequency (Equations~(\ref{eq:one_kick_amp}) and (\ref{eq:K_lm_para})). See Section~\ref{sec:conclusion} and Appendix~\ref{appx:spin_effects} for further discussions.

\subsection{Nonlinear frequency shift}
\label{sec:NL_freq_shift}
In \citet{Yu:21}, we showed that the leading-order nonlinear phase shift can be described by Equation~(\ref{eq:NL_freq_shift_leading_order}).
It is accurate for small $\tilde{E}_a$ yet becomes problematic when $\tilde{E}_a\simeq \left|\omega_a/\Omega\right|$,
as the nonlinear frequency shift can then become greater than the linear frequency, which is unrealistic. Note that $\left(|\omega_a/\Omega|\right)<\tilde{E}_{\rm wb}$, where $\tilde{E}_{\rm wb}\sim 0.1$ is the wave-breaking energy of the mode (the subscript ``wb'' stands for wave-breaking; see Section~\ref{sec:energy_dissipation}). In other words,  Equation~(\ref{eq:NL_freq_shift_leading_order}) becomes inaccurate before the mode breaks by strongly nonlinear effects. 
In order to account for this, we adopt a phenomenological correction to the nonlinear frequency shift 
\begin{equation}
    \frac{\delta \omega_a}{\omega_a} = \frac{\left(\Omega/\omega_a\right) \tilde{E}_a}{1+x|\left(\Omega/\omega_a\right) \tilde{E}_a|},
    \label{eq:NL_freq_shift_phenom}
\end{equation}
where $x>0$ is left as a free parameter. For small $\tilde{E}_a$, this reduces to the leading-order expression derived in \citet{Yu:21}. For large $\tilde{E}_a$, $\delta \omega/\omega_a\to {\rm sign}\left[\Omega\right]/x$, and by choosing $x>1$ we prevent the nonlinear frequency shift from exceeding the original linear frequency. 

Note that the nonlinear frequency shift $\delta \omega_a$ is a function of the mode energy $\tilde{E}_a$, which decays with time between pericenter passages as (see \citealt{Yu:21})
\begin{equation}
    \tilde{E}_a(t) \simeq \frac{\tilde{E}_a^{(0)}}{1 + 2\left[\gamma_a + \Gamma \tilde{E}_a^{(0)}\right]t}\label{eq:E_vs_t}
\end{equation}
(we still approximate $\delta \gamma_a=\Gamma \tilde{E}_a$ for the dissipation and ignore further corrections to it for simplicity).
To account for this, we solve the nonlinear phase as a function of the mode energy $\tilde{E}_a$
\begin{eqnarray}
    \delta \phi_{\rm nl} &=& -\int_{t_0}^{t} \delta \omega_a dt =-\int_{\tilde{E}_a^{(0)}}^{\tilde{E}_a} \frac{\delta \omega}{d\tilde{E}_a/dt} d\tilde{E}_a \nonumber \\
    &=&\frac{\Omega}{2\left(\Gamma - x \gamma |\Omega|/\omega_a\right)}\nonumber \\
    &&\times\log\left\{
    \left[\frac{\gamma + \Gamma \tilde{E}_a}{\gamma + \Gamma \tilde{E}_a^{(0)}}\right]
    \left[\frac{\omega + x |\Omega| \tilde{E}_a^{(0)}}{ \omega + x |\Omega| \tilde{E}_a}\right]
    \right\}.
    \label{eq:dphi_nl_vs_E}
\end{eqnarray}
Note when $x\to0$, the above equation reduces to equation~(41) in \citet{Yu:21}. 

Consequently, the general form (without assuming the energy dissipation is negligible) to evolve the mode when the planet is away from pericenter is given as follows (cf. Equation~(\ref{eq:iter_map_mode_prop})). We first evaluate the magnitude from $|q_{a, k}'^{\rm (0)}|$ to $|q_{a, k}'^{\rm (1)}|$ using Equation~(\ref{eq:E_vs_t}). The phase evolution is then given by $\angle q_{a, k}'^{\rm (1)} = \angle q_{a, k}'^{\rm (0)} + \phi_k$, where $\phi_k = -\omega_a' P_{{\rm orb}, k} + \delta \phi_{\rm nl}$ with $\delta \phi_{\rm nl}$ given by Equation~(\ref{eq:dphi_nl_vs_E}). 

In addition to producing excess phase shift of the f-mode when the planet is away from pericenter, the nonlinear frequency shift may modify the kick the f-mode receives at each pericenter passage. 
This is because the kick depends on the temporal overlap $K_{lm}$ which is a function of the mode frequency (Equation~(\ref{eq:K_lm_para})). 
We account for this effect by replacing $\omega_a'$ with $\omega_a' + \delta \omega_a$ when evaluating $K_{lm}$ in Equation~(\ref{eq:one_kick_amp}). When the mode has built up a significant amount of energy by diffusion, we have $\tilde{E}_{a, k-1}^{(1)}\simeq \tilde{E}_{a, k}^{(0)} \gg \Delta \tilde{E}_{a,k}$. Consequently, we ignore the energy difference before and after the kick and simply use $\tilde{E}_{a, k-1}^{(1)}$ to evaluate a mean frequency shift $\delta \omega_a$. Because $\Omega/\omega_a <0$ (see Table~\ref{tab:models} and \citealt{Yu:21}), we see that including $\delta \omega_a$ makes the kick stronger than the linear case (Equation~(\ref{eq:K_lm_para})). For example, for $\delta \omega_a/\omega_a=-0.1$, the new $K_{22}$ is $80\%$ greater than the linear value.

\subsection{Wave breaking}
\label{sec:energy_dissipation}

While Equation~(\ref{eq:E_vs_t}) describes the damping of the f-mode due to weakly nonlinear interactions, we note that such dissipation may not be sufficient to balance out the energy gain at each pericenter passage. As a result, the mode energy may increase on average as the diffusive process continues. To prevent the f-mode from growing to an unphysically large value, we follow the prescriptions proposed in \citet{Wu:18} (see also \citealt{Vick:19}) and assume an \emph{ad hoc} strongly nonlinear dissipation mechanism in addition to Equation~(\ref{eq:E_vs_t}).\footnote{
Note here we use the word ``weakly nonlinear dissipation'' to refer to the dissipation we calculate from the leading-order nonlinear interaction (i.e., the three-mode interaction) perturbatively. See \citet{Yu:21} for details. This is a first-principle calculation yet it applies only when the mode energy is not too large. 
In contrast, we use the word ``strongly nonlinear dissipation'' to stand for energy dissipation happening when the mode energy growth above the wave-breaking threshold $\tilde{E}_{\rm wb}$. This is a phenomenological model. 
}

As in \citet{Wu:18}, we set $\tilde{E}_{\rm wb}=0.1$ as a threshold energy for the planetary f-mode since it corresponds to the mode having a radial displacement of $\simeq R$ at the surface. If the mode's energy after a pericenter passage exceeds this energy, i.e., $\tilde{E}_{a, k}^{(0)}> \tilde{E}_{\rm wb}$, we then assume the mode breaks through a strongly nonlinear process which removes nearly all but $\tilde{E}_{\rm resi}$ of the mode energy within the next orbital cycle. In other words, if a wave-breaking event happens at the $k$'th passage, we set the mode energy right before the next passage, $\tilde{E}_{a, (k)}^{(1)}$, to the residual value $\tilde{E}_{\rm resi}$. We leave $\tilde{E}_{\rm resi}$ as a free parameter to be explored. The associated nonlinear evolution phase $\delta \phi_{\rm nl}$ is randomly chosen from $[0, 2\pi)$ as such a strongly nonlinear wave-breaking should erase any memory of the original mode phase. 

At this point, we have described all the ingredients necessary to evolve the system over $\sim 10^4\,{\rm yr}$. 
In the next Section, we use them to examine the coupled evolution of the mode and the orbit. 
%We will then examine the coupled evolution of the mode and the orbit Section~\ref{sec:results}.  

% Weakly nonlinear dissipation turned on. 

% When the mode energy grows to a threshold energy, $\tilde{E}_{a, k} > \tilde{E}_{\rm wb}$, we follow the prescription of \citet{Wu:18} and assume a strongly nonlinear process happens. We then remove $99\%$ of the mode energy in the next orbital cycle. We also give a random nonlinear phase shift $\Delta \phi_{\rm nl}\in [0, 2\pi)$ to the f-mode whenever mode breaking happens. 

\section{Coupled evolution of mode and orbit}
\label{sec:results}

Using the formalism described in the previous section, we now solve for the evolution of orbital elements and mode energy during the tidal circularization of a proto-HJ. We consider  two Jupiter-mass models whose parameters are summarized in Table~\ref{tab:models}; we refer to the models according to their radii, namely the $1.1R_J$ model and the $1.7R_j$ model. Both models are constructed using \texttt{MESA} (version 15140; \citealt{Paxton:11, Paxton:13, Paxton:15, Paxton:18, Paxton:19}) and the (linear) eigenmodes are computed using \texttt{GYRE} \citep{Townsend:13, Townsend:18}. The  parameters describing the nonlinear frequency shift $\Omega$ and damping rate $\Gamma$ (Equations~(\ref{eq:NL_freq_shift_leading_order}) and (\ref{eq:NL_damp_shift})) are calculated following the prescription in \citet{Yu:21}. We assume that the dissipation of each daughter mode is due to turbulent convective damping, which we calculate following \citet{Burkart:13}  (see also \citealt{Shiode:12}) and then sum over modes to obtain $\Gamma$.

The $1.1R_J$ model corresponds to an evolved Jupiter with an age $\sim 10^9\,{\rm yr}$. This is the same model as considered in \citet{Yu:21}.\footnote{The values of $\Omega$ and $\Gamma$ presented here are slightly different from the ones in \citet{Yu:21} because different versions of \texttt{MESA} are used to construct the background planet model.} Note that this model has a particularly weak damping due to turbulent convection and therefore the energy is mostly dissipated via strongly nonlinear wave-breaking events as described in Section~\ref{sec:energy_dissipation}. The $1.7R_J$ model corresponds to a relatively young Jupiter with an age of $\sim 10^6\,{\rm yr}$. The key difference is that the younger Jupiter model has a much higher turbulent convective damping rate compared to the more evolved $1.1R_J$ model. As we shall see shortly, the high damping rate could prevent the mode from evolving into the strongly nonlinear regime (at least in the first $\sim 10^4\,{\rm yr}$ of evolution). As a result, we do not necessarily need to invoke our rather ad hoc assumptions about energy dissipation by nonlinear wave-breaking for the $1.7 R_J$ model and its evolution could be  qualitatively different  from that of the $1.1R_J$ model. %It thus lead to qualitatively different evolution of the mode compared to the $1.1R_J$ case. 

\begin{table}[!tb]
\centering
\caption{Properties of the two Jupiter models considered in our study. We write $E_0 {=} GM^2/R {=} E_{0, 43} \times 10^{43}\,{\rm erg}$, $\omega_0{=}\sqrt{GM/R^3}{=}\omega_{0,-4}\times 10^{-4}\,{\rm  rad\,s^{-1}}$.
%as well as $\Gamma = \Gamma_{-10} \times 10^{-10}$. 
%We use the subscript ``prnt'' to denote the parent mode (the prograde f-mode with $l_a=2$) that directly couples to the tide. 
}
\begin{tabular}{ ccccccc } 
 $R$ & $E_{0,43}$ & $\omega_{0, -4}$ & $\omega_{a}$ & $Q_{a}$ & $\Omega$ & $\Gamma$ \\
 \hline
 $1.1R_J$ & $3.1$ & $5.1$ & $1.1\omega_0$ & 0.39 & $-31\,\omega_a$ & $7.2\times10^{-10}\omega_a$ \\
 $1.7R_J$ & $2.0$ & $2.7$ & $1.1\omega_0$ & 0.44 & $-67\,\omega_a$ & $1.1\times10^{-4}\omega_a$
\end{tabular}
\label{tab:models}
\end{table}

\subsection{The $1.1R_J$ model}
\label{sec:results_1p1_Rj}

\begin{figure}
   \centering
   \includegraphics[width=0.45\textwidth]{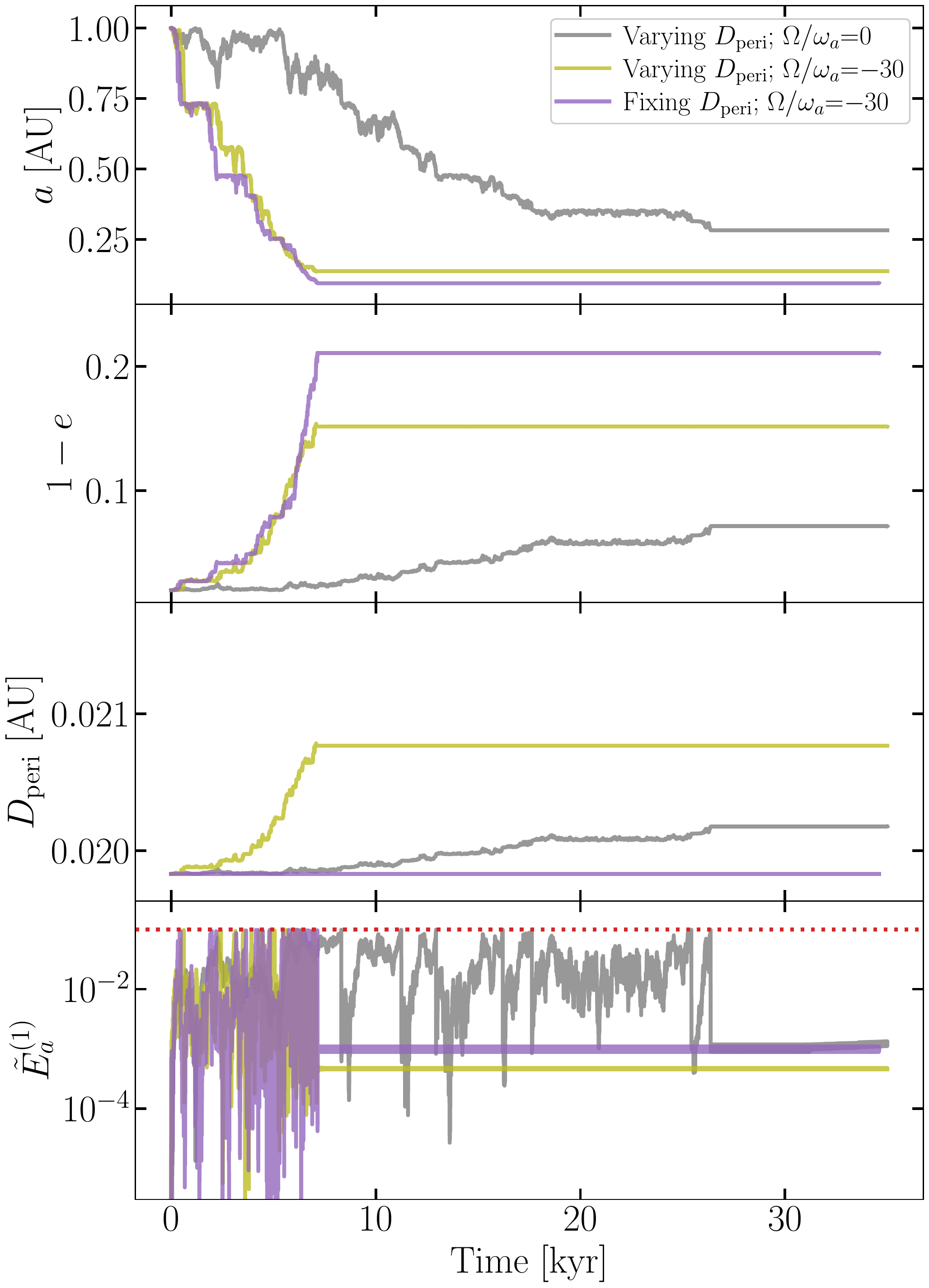} % requires the graphicx package
\caption{Evolution of the orbital elements and mode energy for the $R=1.1R_J$ model over a duration of $\approx 3\times 10^4\,{\rm yr}$. Here the one-kick energy is $\Delta \tilde{E}_1=1.5\times10^{-5}$ and the final orbital period is $P_{\rm orb}^{\rm (f)}=2.8\,{\rm day}$. The gray curves correspond to the linear case where we ignore the nonlinear frequency shifts by setting $\Omega=0$. The olive curves include the nonlinear frequency shift. For comparison, the purple curves show the  trajectories if we prevent the pericenter distance from changing  (and calculate the eccentricity using $e=\left(1-D_{\rm peri}/a\right)$). Because the weakly nonlinear dissipation is small in this model, the mode energy is removed via strongly nonlinear wave breaking whenever it reaches $\tilde{E}_{\rm wb}=0.1$, as indicated by the red-dotted line in the bottom panel. Here we assume $x=3$ and $\tilde{E}_{\rm resi}=10^{-3}$. %\phil{(in the last panel, should the mode energy have subscript k? we've suppressed this in the axis labels for the other panels)}
}
\label{fig:kyr_evol_traj_R_1p1Rj}
\end{figure}

We start by examining evolution trajectories of the evolved (age~$\sim 10^9\,{\rm yr}$) $1.1 R_J$ model.

\subsubsection{Example trajectories with fiducial parameters}

A few representative examples of the orbital evolution are shown in Figure~\ref{fig:kyr_evol_traj_R_1p1Rj}. Here we consider a host star with $M_\ast = 1\,M_\odot$ and an initial orbit with semi-major axis $a_{0}=1\,{\rm AU}$ and  eccentricity $e_{0}=0.980$. The initial pericenter distance is therefore $D_{\rm peri,0}=0.020\,{\rm AU}=3.71 r_{\rm t}$, where the tidal radius $r_{\rm t}\equiv \left(M_\ast/M\right)^{1/3}R$. Assuming constant AM, the final orbital period after complete circularization is $P_{\rm orb}^{(f)}\simeq 2.8\,{\rm day}$. 
We assume that the planet is initially non-rotating and we ignore any subsequent tidal spin-up of the background planet (as discussed in Section~\ref{sec:orbital_AM}, but see Appendix~\ref{appx:spin_effects}). 

We chose the orbital parameters in order that the one kick energy $\Delta \tilde{E}_1$ only slightly exceeds the threshold needed to trigger diffusion~\citep{Yu:21}; its value is $\Delta \tilde{E}_1=1.5\times10^{-5}$. We selected such a $\Delta \tilde{E}_1$ for the following reason.  Before the diffusive tide is triggered, one expects the eccentricity to gradually increase (and hence the pericenter  distance to gradually decrease) due to, e.g., the Lidov-Kozai mechanism~\citep{Wu:03, Fabrycky:07}. Once the the one-kick energy increases above the threshold value, the diffusive tide quickly decouples the planet-host star system from the tertiary perturber and consequently prevents the one-kick energy from increasing further (see, e.g., \citealt{Wu:18, Vick:19}).

In Figure~\ref{fig:kyr_evol_traj_R_1p1Rj}, the gray curves show the linear case in which we ignore the nonlinear frequency shift ($\Omega=0$). For comparison, the olive curves show  the nonlinear case in which the frequency shift  $\Omega/\omega_a=-30$ (Table~\ref{tab:models}). We take a fiducial value of $x=3$ in order to restrict the frequency shift at high energies (see  Equation~(\ref{eq:NL_freq_shift_phenom})). 
By comparing the olive and purple curves, we see the effect of allowing the pericenter distance to vary during the evolution. In particular, the olive curves show the evolution assuming the orbital AM transfer follows the prescription of Section~\ref{sec:orbital_AM}, which then allows us to compute $D_{\rm peri}$ at each pericenter passage from the instantaneous orbital AM. By contrast,  the purple curves assume a constant $D_{\rm peri}$  (as in \citealt{Yu:21}) and the eccentricity is computed using $e_{k} = 1 -\left(D_{\rm peri}/a_{k}\right)$ instead of Equation~(\ref{eq:e_orb_from_L_orb}). Note that for this model, the nonlinear damping $\Gamma$ is too small to balance the energy gain. Thus we have to rely on our ad hoc wave breaking prescription to remove the mode energy (Section~\ref{sec:energy_dissipation}). Here we set $\tilde{E}_{\rm wb}=0.1$ as the wave-breaking threshold and $\tilde{E}_{\rm resi}=10^{-3}$ as the residual after each breaking event. 

From Figure~\ref{fig:kyr_evol_traj_R_1p1Rj}, we see that once the diffusive tide is triggered, it can reduce the orbital semi-major axis by nearly an order of magnitude over a timescale of only $\approx 10^4\,{\rm yr}$. 

To help relate our results to  observations, we can express them in terms of an effective tidal quality factor $\mathcal{Q}$, defined such that~\citep{Goldreich:89} %\phil{(Should we use the $\dot{e}$ expression here instead of the $\dot{a}$ expression?)}
\begin{equation}
    \frac{1}{a}\frac{da}{dt} = -\frac{21}{64}\Omega_{\rm orb}\frac{M_\ast}{M}\frac{aR^5}{D_{\rm peri}^6}\frac{k_2}{\mathcal{Q}},
\end{equation}
where $k_2$ is the tidal Love number, whose typical value is $0.3\lesssim k_2 \lesssim 0.5$. On average, we find that the orbit decays by about $1\,{\rm AU}$ in $10^4\,{\rm yr}$ (Figure~\ref{fig:kyr_evol_traj_R_1p1Rj}), which implies
\begin{eqnarray}
    \frac{\mathcal{Q}}{k_2}\simeq 14 
    \left(\frac{P_{\rm orb}}{1\,{\rm yr}}\right)^{1/3}
    \left(\frac{D_{\rm peri}}{0.02\,{\rm AU}}\right)^{-6}
    \left(\frac{|da/dt|}{1\,{\rm AU}/10^4\,{\rm yr}}\right)^{-1}.
    \label{eq:tidal_Q_factor}
\end{eqnarray}
This is about 4 to 5 orders of magnitude smaller than the empirically determined $\mathcal{Q}$ inferred for the Jupiter-Io system. 
The corresponding tidal lag time \citep{Ogilvie:14} is $\tau\equiv 1/(\mathcal{Q}\omega_{\rm d})\simeq 10^3\,{\rm s}$ at a driving frequency $\omega_{\rm d}=2\pi/(1\,{\rm day})$, which motivates our choice of $\tau$ in Equation~(\ref{eq:tau_tidal_lag}).

It is important to note that the diffusive tide does not circularize the orbit down to zero eccentricity.  Instead, in all three cases plotted in Figure~\ref{fig:kyr_evol_traj_R_1p1Rj}, the evolution due to diffusive tide terminates when the eccentricity is still high. Under the linear theory (gray curves), the evolution terminates at the largest orbital eccentricity and we still have $e>0.9$ when the diffusive evolution stalls. Including the nonlinear frequency shift helps circularizing the orbit further, though we still have $e\sim 0.8$ when the diffusive tide terminates. Taking into account the increase in $D_{\rm peri}$ also hinders the circularization process compared to the case where $D_{\rm peri}$ is held constant.

\subsubsection{Statistics of the termination point}

Because the diffusive tidal evolution is a stochastic process, the termination point of a particular system, as shown in Figure~\ref{fig:kyr_evol_traj_R_1p1Rj}, is sensitive to the initial conditions.  Here we examine the statistics of the termination point in Figure~\ref{fig:terminal_pt_strongly_damped} by giving the mode a small initial amplitude $|q_{a,0}|=10^{-5}$ but a random phase. Because the evolution is chaotic~\citep{Vick:18}, varying the initial conditions allows us to obtain different trajectories for the same set of parameters. The median values of the distribution of the termination point are connected with solid lines in Figure~\ref{fig:terminal_pt_strongly_damped}, and each error bar corresponds to the 20'th and 80'th percentiles for the specific set of parameters. If a parameter is not explicitly listed in the legend, its value is that given in Figure~\ref{fig:kyr_evol_traj_R_1p1Rj}. 

Along the x-axis, we vary the value of the nonlinear frequency shift $\Omega/\omega_a$; the leftmost point (with $\Omega=0$) corresponds to the linear case. i.e., ignoring nonlinear mode interactions. Towards the right, the magnitude of the nonlinear frequency (and hence phase) shift increases, which helps prolong the diffusive tidal evolution and drive the system to smaller terminal values of $a$ and $e$. Specifically, when $\Omega=0$ and we ignore the nonlinear frequency shift, the evolution terminates at $e> 0.9$  under all the scenarios we consider in Figure~\ref{fig:terminal_pt_strongly_damped}. When we include the nonlinear frequency shift, the circularization continues down  to $e\simeq 0.8-0.85$. 

The effect of allowing the pericenter distance to vary can be seen by comparing the gray and olive curves. For the linear case, it does not typically affect the results since the evolution stalls before the pericenter can change significantly. By contrast, when the nonlinear frequency shift is incorporated, the circularization continues further and the increasing pericenter distance plays a more important role in terminating the orbital evolution and preventing the orbital eccentricity from dropping below $e\simeq 0.8$. 

Another factor limiting the circularization is the value of $x$, introduced to avoid the nonlinear frequency shift from becoming greater than the original linear eigenfrequency (Section~\ref{sec:NL_freq_shift}). Its effect can be seen by comparing the olive and purple curves. As $x$ increases, the maximum fractional frequency shift ($\simeq 1/x$) decreases in our phenomenological model  (Equation~(\ref{eq:NL_freq_shift_phenom})). This causes the circularization to stall at greater values of $e$ as $x$ increases. It also causes the plateau of both $a$ and $e$ for large values of $|\Omega/\omega_a|$ in Figure~\ref{fig:terminal_pt_strongly_damped}.%\nevin{I'm not sure I completely understand the difference between this last sentence and the second-to last sentence.} 

Lastly, we examine the effect of $\tilde{E}_{\rm resi}$, the residual energy after a wave-breaking event, by comparing the olive and cyan curves. While the cyan curve has an $\tilde{E}_{\rm resi}$ that is 50 times greater than the one used for the olive curve, its terminal $a_{\rm orb}$ is only slightly smaller. Consequently, the circularization is not sensitive to the value of $\tilde{E}_{\rm resi}$ (to understand why, see discussions following Equations~(\ref{eq:a_orb_term_br}) and (\ref{eq:a_orb_term_nl}) below). 

\subsubsection{Analytic estimates}

\begin{figure}
   \centering
   \includegraphics[width=0.45\textwidth]{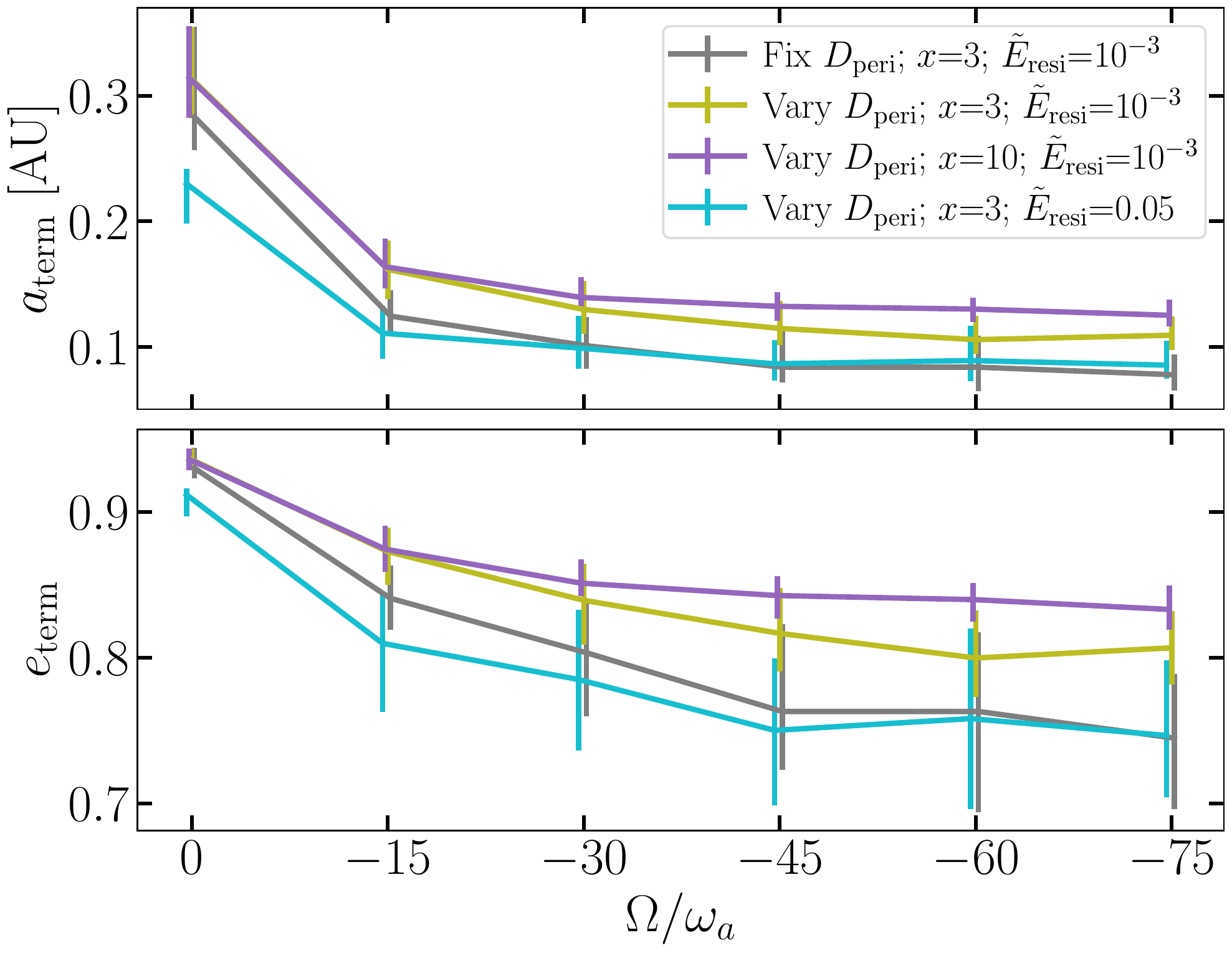} % requires the graphicx package
\caption{Semi-major axis and eccentricity at which the diffusive tide terminates % for the $1.1R_J$ model. Here we fix the one-kick energy to be $\Delta \tilde{E}_1=1.5\times10^{-5}$ and the wave-breaking threshold $\tilde{E}_{\rm wb}=0.1$. 
%At each value of $\Omega/\omega_a$, we run 100 different realizations of the evolution by randomly selecting the f-mode's initial phase (its initial magnitude is fixed at $|q_0|=10^{-5}$). 
in the $1.1R_J$ model assuming an orbit with $\Delta \tilde{E}_1=1.5\times10^{-5}$ and $P_{\rm orb}^{\rm (f)}=2.8\,{\rm day}$. The error bars account for differences in the initial conditions (see text). The central point (error bar) corresponds to the median (20 and 80 percentiles) of the resultant distribution. 
If we ignore the nonlinear phase shift ($\Omega/\omega_a=0$), $e_{\rm term}\simeq 0.95$ whereas when we include it $e_{\rm term}\simeq 0.8$. %\nevin{Should the legend say $E_{\rm wb}$ not $E_{\rm th}$?  Also, in the text we give  $\bar{E}_{\rm resi}$ not $\bar{E}_{\rm resi}/\bar{E}_{\rm th}$.  To be consistent, maybe give the former in the legend?} 
%When $D_{\rm peri}$ varies (it increases as the orbit evolves), the termination happens earlier (at a higher eccentricity). Smaller $x$ [larger saturation frequency; Equation~(\ref{eq:NL_freq_shift_phenom}); purple vs. olive curves] or greater $\tilde{E}_{\rm resi}$ (cyan vs. olive curves) maintains the evolution further to smaller eccentricities. 
}
\label{fig:terminal_pt_strongly_damped}
\end{figure}

The numerical results shown in Figures~\ref{fig:kyr_evol_traj_R_1p1Rj} and \ref{fig:terminal_pt_strongly_damped} can be understood analytically. The key is that for the diffusive tide to remain active, the phase difference between two adjacent orbits,\footnote{Consistent with the notation in \citet{Yu:21}, we use the symbol $\Delta$ to specifically indicate the difference between two adjacent orbit. } $\Delta \phi_k=\phi_k - \phi_{k-1}$,  needs to satisfy $|\Delta \phi_k| \gtrsim 1\,{\rm rad}$ on average (see also the discussions in \citealt{Wu:18, Vick:19, Yu:21}). By evaluating $\Delta \phi_k$ as a function of, e.g., the semi-major axis of the orbit $a$, and then inverting the relation, we can estimate the termination point when the phase difference is no longer large enough to maintain the diffusion. 

Under the linear theory, we have
\begin{equation}
    \Delta \phi_{\rm br, k} = -\omega_a' \Delta P_{\rm orb, k} =\frac{3}{2}\omega_a' P_{\rm orb, k}\frac{\Delta \tilde{E}_k}{\big{|}\tilde{E}_{\rm orb, k}\big{|}},
    \label{eq:dphi_br_k}
\end{equation}
where the subscript ``br'' stands for back-reaction, as it is the only channel to create $\Delta \phi_k$ under the linear theory. 
To evaluate the energy change $\Delta \tilde{E}_k$, we use Equation~(\ref{eq:Delta_E_k}). The evolution terminates when $|\Delta \phi_{\rm br, k}|\sim 1\,{\rm rad}$ (but not exactly 1\,rad). Since we are mostly interested in how the final $a$ scales with different parameters, and our estimates of both $\Delta \tilde{E}_{a,k}$ and the terminal $|\Delta \phi_{\rm br, k}|$ are accurate only to order unity,  we drop all the numerical prefactors and simply absorb them into an overall scaling factor $\lambda_{\rm br}$. 

Using this approach,  the semi-major axis when diffusive growth terminates is given by
\begin{eqnarray}
    &&a^{\rm (br)}_{\rm term} \simeq 0.5\,{\rm AU} 
    \nonumber \\
    &&\times
    \left(\frac{\lambda_{\rm br}}{0.1}\right)^{2/5}
    \left(\frac{\Delta \tilde{E}_1}{10^{-5}}\right)^{-1/5}
    \left(\frac{\tilde{E}_{a,k}}{10^{-3}}\right)^{-1/5}
    \nonumber \\
    &&\times 
    \left(\frac{\omega_a'}{\omega_J}\right)^{-2/5}
    \left(\frac{M}{M_J}\right)^{-2/5}
    \left(\frac{R}{R_J}\right)^{2/5}
    \left(\frac{M_\ast}{M_\odot}\right)^{3/5},
    \label{eq:a_orb_term_br}
\end{eqnarray}
where $\omega_J^2 \equiv GM_J/R_J^3$. It is interesting to note that $a^{\rm (br)}_{\rm term}$ has a weak dependence on $\tilde{E}_{a,k}$($\simeq \tilde{E}_{\rm resi}$, since it  typically terminates after a wave-breaking event, when the mode energy is small). This is why in Figure~\ref{fig:terminal_pt_strongly_damped} the results depend weakly on $\tilde{E}_{\rm resi}$. The dependence on $\Delta \tilde{E}_{1}$ is also weak, although it is partially compensated by the fact that $\Delta \tilde{E}_1$ has a very sharp dependence on $D_{\rm peri}$, scaling as $\Delta \tilde{E}_1\propto D_{\rm peri}^{-20}$ (Equations~(\ref{eq:one_kick_amp}) and (\ref{eq:K_lm_para}); see also \citealt{Wu:18}). Consequently, it requires a $\simeq 10\%$ change in $D_{\rm peri}$ in order to change $a_{\rm term}^{\rm (br)}$ by about $50\%$. From Figure~\ref{fig:kyr_evol_traj_R_1p1Rj}, we see that the change in $D_{\rm peri}$ is much less than $10\%$ for $\tilde{E}_{\rm resi}=10^{-3}$. This is why the gray and olive bars at $\Omega/\omega_a=0$ largely overlap in Figure~\ref{fig:terminal_pt_strongly_damped}. 
In other words, the evolution is terminated mostly due to the decrease of $a$, which reduces both the time $P_{\rm orb}$ for the phase to accumulate and the fractional change $\Delta P_{\rm orb}/P_{\rm orb}$ (as the orbit becomes more tightly bound; see also the discussion in \citealt{Yu:21}). The increase of $D_{\rm peri}$ is a subdominant effect in this case. 
For greater values of $\tilde{E}_{\rm resi}=0.05$, the circularization continues further and $D_{\rm peri}$ could change by a few percent. The increase in $D_{\rm peri}$ now starts to play a role (though still mild) and as a result the final $a$ shown in the cyan bar in Figure~\ref{fig:terminal_pt_strongly_damped} is less than a factor of $50^{1/5}\simeq 2$ smaller than the gray one.

For typical propto-HJs, we expect the phase shift to be dominated by the nonlinear frequency shift \citep{Yu:21}.  In that case,  it is given by \footnote{While this effect originates from nonlinear mode interactions, we note that the phase shift it induces scales with the change in mode energy as $\Delta \phi_{{\rm nl}, k}\propto \Delta \tilde{E}_{a,k}$. This is the same scaling as the phase shift due to the tidal back-reaction under the linear theory (Equation~(\ref{eq:dphi_br_k})). Therefore the two effects are formally the same order in $\Delta \tilde{E}_{a,k}$. }
\begin{equation}
    \Delta \phi_{{\rm nl}, k} \simeq -\omega_a \left(\frac{\Omega}{\omega_a}\right) P_{{\rm orb}, k} \Delta \tilde{E}_k,
    \label{eq:dphi_nl_k}
\end{equation}
where for simplicity we used the leading-order nonlinear frequency shift (Equation~(\ref{eq:NL_freq_shift_leading_order})) and approximated $\delta \phi_{\rm nl}\simeq -\Omega \tilde{E}_{a,k}P_{{\rm orb},k}$ for each orbit (which applies when $\gamma P_{{\rm orb},k} {\ll} 1$ and $\Gamma \tilde{E}_{a,k}P_{{\rm orb},k} {\ll} 1$). The termination point of $a$ can again be found by setting $|\Delta \phi_{\rm nl, k}| \sim 1\,{\rm rad}$, which gives
\begin{eqnarray}
    &&a^{\rm (nl)}_{\rm term} \simeq 0.1\,{\rm AU} 
    \nonumber \\
    &&\times 
    \left(\frac{\lambda_{\rm nl}}{0.3}\right)^{2/3}
    \left(\frac{\Delta \tilde{E}_1}{10^{-5}}\right)^{-1/3}
    \left(\frac{\tilde{E}_{a,k}}{10^{-3}}\right)^{-1/3}
    \nonumber \\
    &&\times 
    \left(\frac{\omega_a}{\omega_J}\right)^{-2/3}
    \left(\frac{|\Omega|/\omega_a}{30}\right)^{-2/3}
    \left(\frac{M_\ast}{M_\odot}\right)^{1/3},
    \label{eq:a_orb_term_nl}
\end{eqnarray}
where we have absorbed the uncertainties in $\Delta \tilde{E}_{a, k}$ and $\Delta \phi_{{\rm nl}, k}$ together with all the order unity factors into an overall scaling factor $\lambda_{\rm nl}$. 
%Although nonlinear interactions drive the diffusive tidal evolution to smaller $e$ and $a$ compared to the linear case (Equation~(\ref{eq:a_orb_term_br}) and Figures~\ref{fig:kyr_evol_traj_R_1p1Rj} and \ref{fig:terminal_pt_strongly_damped}), they are still insufficient to fully circularize the orbit.\nevin{ Do we want to say this again here? It breaks the flow of the paragraph a bit and we say it earlier in the paragraph starting around line 592.} 
Note that  Equation~(\ref{eq:a_orb_term_nl}) does not account for the reduction of the nonlinear frequency shift (Equation~(\ref{eq:NL_freq_shift_phenom})), which is why in Figure~\ref{fig:terminal_pt_strongly_damped} the terminal $a$ decreases  slower than $|\Omega/\omega_a|^{-2/3}$. 

Comparing Equations~(\ref{eq:a_orb_term_br}) and (\ref{eq:a_orb_term_nl}), we see that the termination point depends more sensitively on $D_{\rm peri}$ in the nonlinear case.  Specifically, $a^{\rm (br)}_{\rm term} \propto \Delta \tilde{E}_1^{-1/5}\propto D_{\rm peri}^{4}$ compared to $a^{\rm (nl)}_{\rm term} \propto \Delta \tilde{E}_1^{-1/3}\propto D_{\rm peri}^{6.7}$. For example, a $5\%$ change in $D_{\rm peri}$ can modify $a^{\rm (nl)}_{\rm term}$ by about $40\%$. Due to the increase in  $D_{\rm peri}$ with time, the circularization stalls at $e\gtrsim 0.7$ even under the most favorable conditions (i.e., small $x$, large $\tilde{E}_{\rm resi}$, and large $|\Omega/\omega_a|$).

% Figure~\ref{fig:terminal_pt_strongly_damped} shows the statistical properties of the termination point. 

\subsection{The $R=1.7R_J$ model}
\label{sec:results_1p7_Rj}

\begin{figure}
   \centering
   \includegraphics[width=0.45\textwidth]{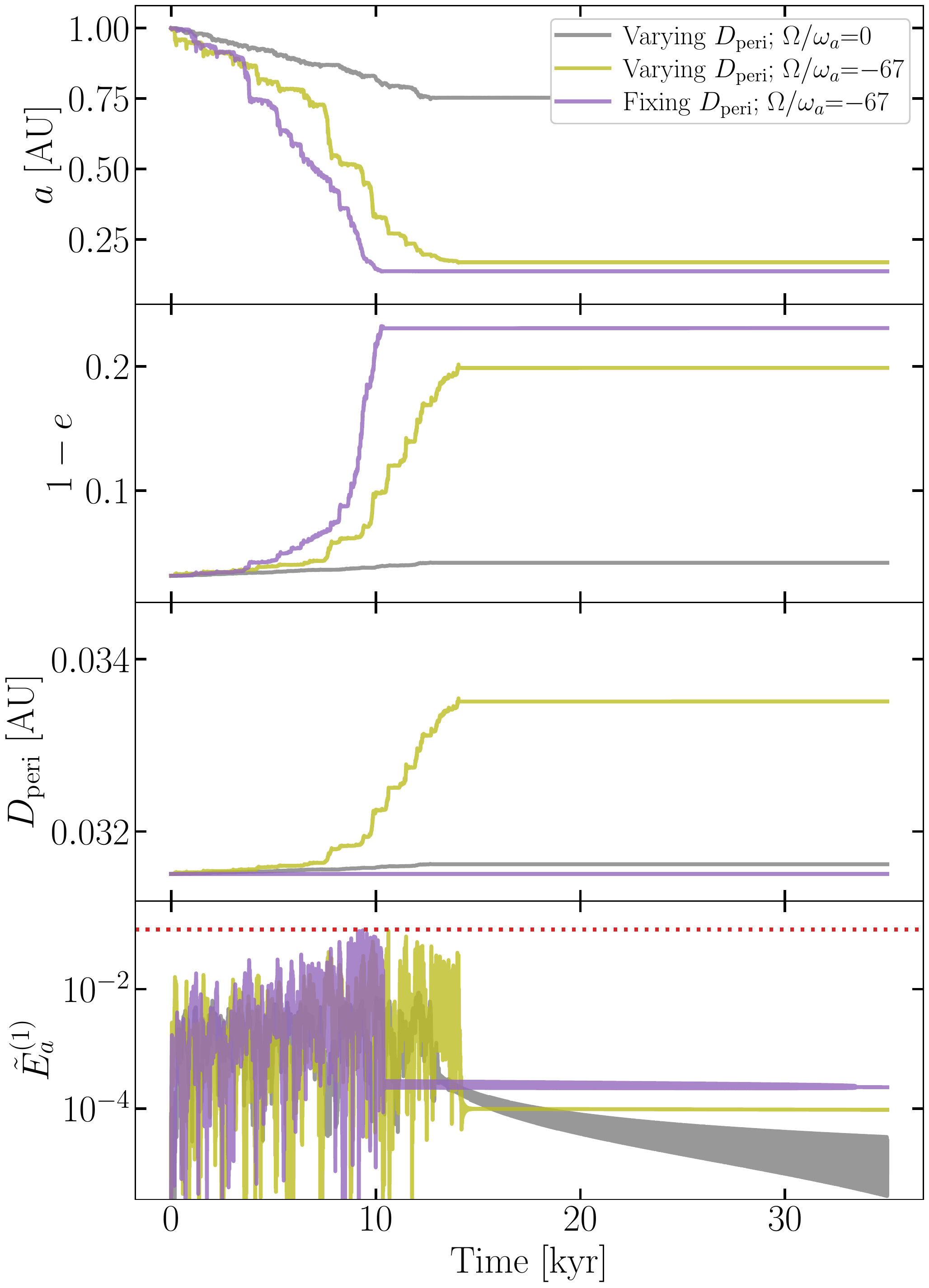} % requires the graphicx package
\caption{Evolution trajectories of the $1.7R_J$ model. Here we choose a pericenter distance of $D_{\rm peri}=0.0315\,{\rm AU}$ so that the one-kick energy is $\Delta \tilde{E}_1=1.00\times10^{-5}$ and the final orbital period is $P_{\rm orb}^{\rm (f)}=5.6\,{\rm day}$. 
Note that during the first $10^4\,{\rm yr}$ of the evolution, the mode energy is below the wave-breaking limit (red-dotted line in the forth panel) as the dissipation due to turbulent convection and nonlinear interaction is sufficient to balance the energy gain from the orbit. 
%The key difference from Figure~\ref{fig:kyr_evol_traj_R_1p1Rj} is that at least in the first 10\,kyr of the evolution, the weakly nonlinear dissipation is sufficient to damp out the energy gain at each pericenter passage, and therefore the mode evolves in a more regulated way with the mode energy below the wave-breaking threshold, $\tilde{E}_a<\tilde{E}_{\rm wb}$. Eventually, wave-breaking still happens after $\simeq 10\,{\rm kyr}$ as the orbital period becomes short and there is not enough time for the weakly nonlinear dissipation to damp out the energy gain in one orbital cycle.   
}
\label{fig:kyr_evol_traj_R_1p7Rj}
\end{figure}

We now discuss the results of the $1.7R_J$ model.  
Notably, its $\Gamma$ is more than five orders of magnitude larger than that of the $1.1R_J$ model (Table \ref{tab:models}). We will see that the f-mode does not undergo  wave-breaking as  a result.

In Figure~\ref{fig:kyr_evol_traj_R_1p7Rj} we show the evolution of the orbital elements and mode energy assuming representative parameter values. We again assume that the initial orbital semi-major axis is $a_{\rm orb}=1\,{\rm AU}$.  However,  we change the pericenter distance to $D_{\rm peri}=0.0315\,{\rm AU}\simeq3.82\,r_{\rm t}$ in order that the one-kick energy $\Delta \tilde{E}_1=1.00\times10^{-5}$ is again only slightly above the threshold needed to trigger diffusion. The final orbital period in this case is $P_{\rm orb}^{\rm (f)}=5.6\,{\rm day}$. Other parameters used to generate the plot are $(x, \tilde{E}_{\rm wb}, \tilde{E}_{\rm resi})=(3, 0.1, 10^{-3})$. 

The evolution of $\tilde{E}^{(1)}_{a,k}$  in the $1.7R_J$ model (bottom panel in Figure~\ref{fig:kyr_evol_traj_R_1p7Rj}) shows qualitatively different behavior than in the $1.1R_J$ model. Whereas for the $1.1R_J$ model, $\tilde{E}^{(1)}_{a,k}$ is driven to the wave-breaking value $\tilde{E}_{\rm wb}$ (red-dotted line) in about a hundred years and reaches $\tilde{E}_{\rm wb}$  repeatedly in the subsequent evolution, for the $1.7R_J$ model, it stays below $\tilde{E}_{\rm wb}$ for at least the first $10^4$ yr of the evolution when $a\gtrsim 0.2\,{\rm AU}$. This is because the $1.7R_J$ model has a much larger $\Gamma$ and therefore the nonlinear mode interactions are sufficiently dissipative (Equation~(\ref{eq:E_vs_t})) that they can balance the energy gained by the mode at each pericenter passage. 

We can estimate the steady-state energy of the mode based on the  energy balancing argument %\nevin{Why is it $\tilde{E}^{(0)}_{a,k}$ here and below but $\tilde{E}^{(1)}_{a,k}$ previously?}
\begin{equation}
    2\Gamma \tilde{E}_{a, k}^{(0)} P_{\rm orb} \simeq \Delta \tilde{E}_{a, k}\sim \sqrt{\tilde{E}_{a, k}^{(0)} \Delta \tilde{E}_1 },
\end{equation}
where we have used $\tilde{E}_{a, k}^{(1)}\simeq \tilde{E}_{a, k}^{(0)}\gg \Delta \tilde{E}_{a, k}\gg \Delta \tilde{E}_{1}$ when the mode has built up its energy via the diffusive growth.
Dropping factors of order unity, we have
\begin{eqnarray}
    &&\tilde{E}_{a, k}^{(0)}\simeq 10^{-4}
    \left(\frac{\Delta \tilde{E}_1}{10^{-5}}\right)
    \left(\frac{\Gamma/\omega_a}{10^{-4}}\right)^{-2}
    \left(\frac{\omega_a}{\omega_J}\right)^{-2} 
    \nonumber \\
    && \times 
    \left(\frac{M_\ast}{M_\odot}\right)
    \left(\frac{a_{k}}{{\rm AU}}\right)^{-3}.
\end{eqnarray}
The energy dissipation mechanism is therefore based on a first-principle calculation  involving turbulent convection and leading-order nonlinear interactions rather than the ad hoc model  of wave breaking used in the $1.1R_{J}$ case (at least for the first $\sim 10^4\,{\rm yr}$ of the evolution).  
Note, however, that $\tilde{E}_{a, k}^{(0)}\propto a^{-3}$ because as the orbit shrinks, there is less time for the dissipation to take place. As $a$ decays from $1\,{\rm AU}$ to about $0.2\,{\rm AU}$, the steady-state $\tilde{E}_{a, k}^{(0)}$ increases by about a factor of 100, and strongly nonlinear wave-breaking may still eventually occur. 

Although the mode now has a qualitatively different evolution trajectory, the orbital elements nonetheless evolve in similar ways for the $1.7R_J$ and $1.1R_J$ models. In particular, the diffusive tide drives significant orbital decay within  about $10^4$\,yr, leading to an effective quality factor (or tidal lag time) similar to the $1.1R_J$ model (see  Equation~(\ref{eq:tidal_Q_factor})). In addition, the circularization again stalls when the eccentricity is still relatively high ($e\gtrsim 0.8$).

\begin{figure}
   \centering
   \includegraphics[width=0.45\textwidth]{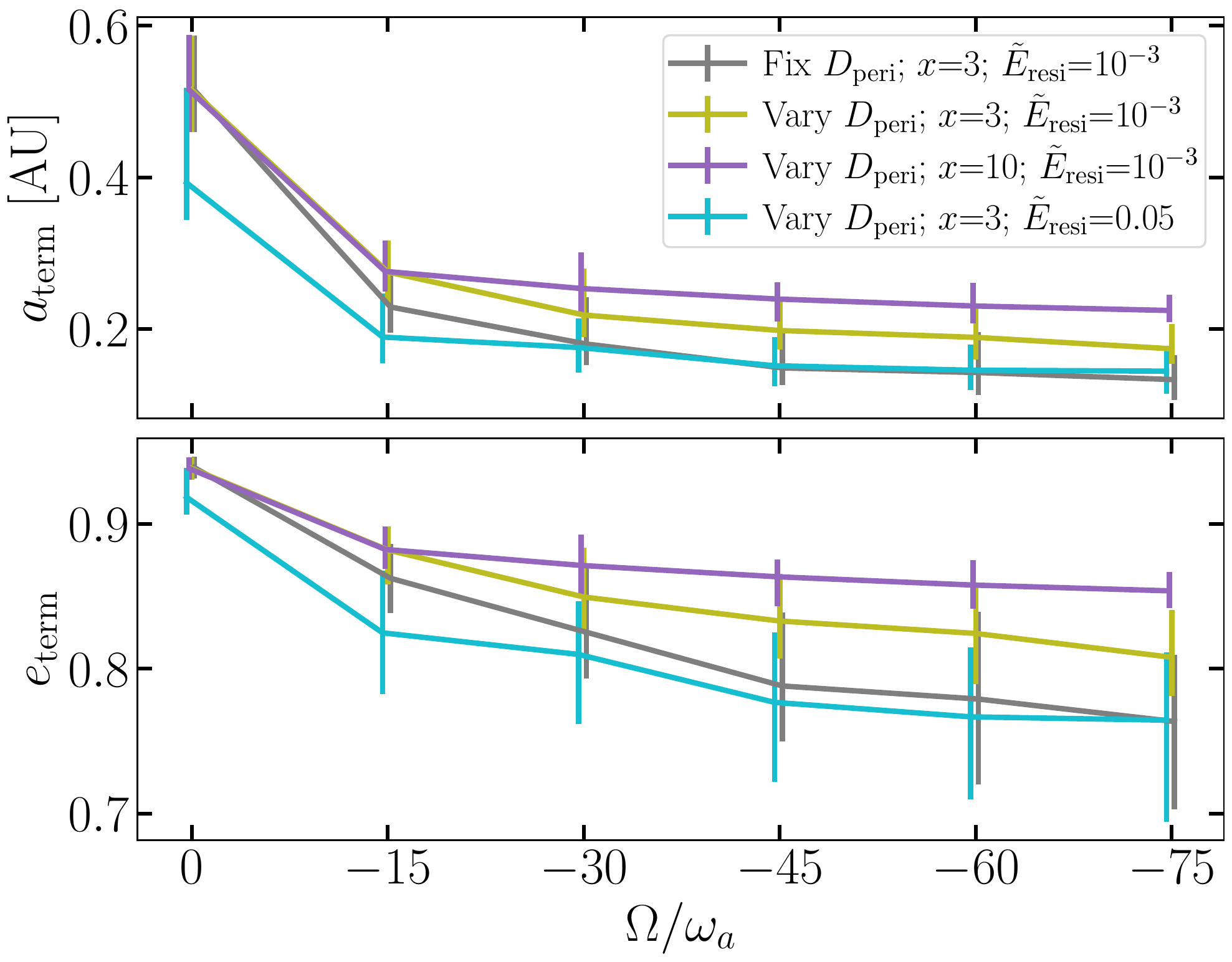} % requires the graphicx package
\caption{Similar to Figure~\ref{fig:terminal_pt_strongly_damped} but for the $1.7R_J$ model with one-kick energy $\Delta \tilde{E}_1=10^{-5}$ and final orbital period $P_{\rm orb}^{(f)}=5.6\,{\rm day}$. }
\label{fig:terminal_pt_weakly_damped}
\end{figure}

In Figure~\ref{fig:terminal_pt_weakly_damped}, we examine the statistical properties of the termination point for the $1.7R_J$ model. The conclusions reached for the $1.1R_J$ case (Figure~\ref{fig:terminal_pt_strongly_damped}) apply again here, including the fact that the terminal eccentricity under the linear theory is $e_{\rm term} > 0.9$ whereas $e_{\rm term}\simeq 0.8$ when nonlinear effects are accounted for.

\section{Summary and Discussion}
\label{sec:conclusion}

In this work, we revisited the eccentricity distribution of migrating proto-HJs under the influence of the diffusive tide~(Section \ref{sec:number_along_track}). We showed that the diffusive growth of the f-mode results in an especially large tidal lag time (small quality factor; Equations~(\ref{eq:tau_tidal_lag}) and (\ref{eq:tidal_Q_factor})) that quickly decreases the initially high eccentricity ($e\simeq 1$) to more moderate values ($e\simeq 0.8-0.95$).  We found that this can explain the observed paucity of super-eccentric proto-HJs with $e>0.9$ \citep{Dawson:15}, as also found in the linear-theory based diffusive tide study by \citet{Wu:18}. %\nevin{Maybe end the paragraph here and don't need to say next two sentences?}Even if the diffusion terminates at a relatively high eccentricity of $e_{\rm term}=0.95$ (gray curve in Figure~\ref{fig:num_vs_e}), we estimated that the expected number of super-eccentric proto-HJ in \emph{Kepler}'s data should be less than 1. Our study thus explains the observed paucity of super-eccentric Jupiters~\citep{Dawson:15}. 

%To further understand the details of the diffusive evolution, it would rely on the population of highly-eccentric proto-HJs with $0.7<e<0.9$. The current observational constraints are not well established for this population because such systems are hard to be identified using the photo-eccentric effect~\citep{Dawson:15}. Our study thus serves as a motivation for the development of novel searching techniques and strategies for highly-eccentric population.

Detecting proto-HJs with $0.8<e < 0.9$ would more fully test the viability of high-eccentricity migration under the action of the diffusive tide. 
% Although detecting such systems was challenging with \emph{Kepler}  using the photo-eccentric effect~\citep{Dawson:15},\nevin{this makes it sounds like the photo-eccentric effect is the only way to detect them with Kepler.  Is that right?} the \emph{TESS} mission may start to place interesting constraints on such a population.\nevin{Maybe say why TESS can do it but not Kepler?} Indeed, 
The recently discovered proto-HJ TOI-3362b by \emph{TESS}~\citep{Dong:21} with $e= 0.815$ is consistent with a planet that has ended the rapid phase of diffusive evolution and entered the  slow circularization phase  (Equation~(\ref{eq:tau_tidal_lag})). The  discovery of additional systems like TOI-3362b  should help test the diffusive evolution scenario and constrain the termination eccentricity $e_{\rm term}$. 

% \hang{On the other hand, the system HD 80606b~\citep{Wu:03} has an eccentricity of $e=0.983$, yet its radius is too small to trigger diffusion given its pericenter distance~\citep{Wu:18}. Therefore, the analysis presented in Section~\ref{sec:number_along_track}, while extending the original study by \citet{Socrates:12}, still has limitations as the eccentricity distribution depends on not only the orbital AM $L$ but also on parameters like the planetary radius, etc. More sophisticated examination of this problem will thus be desired.}

%\nevin{Does the following capture everything of the previous paragraph?} 
The system HD 80606b~($e=0.983$; \citealt{Wu:03}) might seem to be an obvious candidate for a proto-HJ undergoing diffusive evolution.  However, its radius is too small to trigger diffusion given its pericenter distance~(\citealt{Wu:18}; see also Appendix~\ref{appx:P_f_vs_R}).  This nonetheless raises an important point regarding the analysis presented in Section~\ref{sec:number_along_track}. Namely, the triggering and termination of the diffusive tide (as captured by Equation~(\ref{eq:tau_tidal_lag}))  is a complicated function of the orbital and planetary parameters and care must be taken when comparing observations of individual system with the theory.

The phenomenological model in Section~\ref{sec:number_along_track} was motivated by our  analysis of the high-eccentricity evolution of proto-HJs over $\mathcal{O}(10^4)\,{\rm yr}$, which we presented in Sections~\ref{sec:formalism} and~\ref{sec:results}.  This  time span is  long enough to cover the diffusive tidal evolution from its onset to termination. 
We extended the treatment of our previous work, \citet{Yu:21}, in the following three ways.  First, we incorporated into the iterative mapping formalism the evolution of the orbital AM and hence the pericenter distance (Section~\ref{sec:orbital_AM}).  Second, we introduced a phenomenological model (Equation~(\ref{eq:NL_freq_shift_phenom})) that prevents  unphysically large nonlinear frequency shifts when $|\Omega|\tilde{E}_{a, k}\sim \omega_a$. 
%To avoid an unphysically large nonlinear frequency shift when $|\Omega|\tilde{E}_{a, k}\sim \omega_a$, we proposed a phenomenological model in Equation~(\ref{eq:NL_freq_shift_phenom}). 
%The effects of the nonlinear frequency shift on the tidal kick were considered in Section~\ref{sec:NL_freq_shift}. 
Lastly,  we included weakly nonlinear energy dissipation due to  turbulent convective damping and leading-order nonlinear mode interactions, as well as a phenomenological model for strongly nonlinear energy dissipation in the event of wave-breaking of the f-mode (Section~\ref{sec:energy_dissipation}). 

With these new ingredients, we examined the coupled mode-orbit evolution for two $M=M_J$ planetary models, one with radius $1.1R_J$ and the other with radius $1.7 R_J$.  These were meant to represent  an old and young Jupiter, respectively (approximately $10^9$ and $10^6$ years old).  The daughter modes in the $1.1 R_J$ model had an especially small turbulent damping rate.  Its f-mode therefore  grew diffusively until it reached wave-breaking, resulting in  episodes of extreme, sudden energy dissipation (Section~\ref{sec:results_1p1_Rj}).  By contrast, the turbulent damping rates of the daughter modes in the $1.7 R_J$ model were orders of magnitude larger than in the $1.1 R_J$ model.  As a result, its f-mode achieved  a steady state in which  the energy gained at each pericenter passage was balanced by weakly nonlinear dissipation from the excited daughters.  We therefore did not need to invoke our phenomenological treatment of wave breaking for the $1.7 R_J$ model; the analysis instead relied entirely on first-principle calculations (at least for the first $10^4\,{\rm yr}$ of the evolution when $a>0.2\,{\rm AU}$; Section~\ref{sec:results_1p7_Rj}). 

Despite these differences, we  found that the two models undergo similar orbital evolution under the diffusive tide. Specifically, the orbital semi-major axis decreases by a factor of a few to ten in $\simeq 10^4\,{\rm yr}$. If we include (neglect) nonlinear mode interactions, the eccentricity decreases from $e\simeq 1$ to  $e\simeq 0.8$ ($e \gtrsim 0.9$) before the diffusive growth of the f-mode terminates.  Although the impact of nonlinear mode interactions on the termination eccentricity may seem small, we showed in Section~\ref{sec:number_along_track} that it translates to a significant difference in the number of high-$e$ proto-HJs expected to be found in surveys like \emph{Kepler}.

%however, it could not circularize the orbit all the way to a nearly zero eccentricity. Instead, we found that under the linear theory, the diffusive evolution stalls when the eccentricity is still high, $e>0.9$. Including nonlinear mode interactions allows the orbit to circularize further to a smaller yet still significant eccentricity of $e\simeq 0.8$. 

Although our work can explain the paucity of proto-HJs with $e>0.9$, our understanding of high-eccentricity tidal migration under the current framework is still far from complete.  An especially important issue is the nature of the mechanism that  drives the subsequent evolution towards $e\simeq 0$ after the diffusive tide terminates. Even under favorable parameters and assuming the planet can efficiently get rid of the excess AM, we still find significant  eccentricities of $e\simeq 0.8$ at termination. 
In Section~\ref{sec:number_along_track}, we  assumed a model with a constant tidal lag time for post-diffusive tidal evolution (as \citealt{Socrates:12} did).  \citet{Vick:19} showed that the subsequent circularization can be achieved within $10^9\,{\rm yr}$ if $\tau\gtrsim 1\,{\rm s}$.  However, first-principle calculations predict a lag time that is orders of magnitude smaller and closer to $\tau\sim 10^{-8}\,{\rm s}$  \citep{Goldreich:77}.  One possibility is that the damping rate in the planet due to convection is much larger than the original calculations by \cite{Goldreich:77} found (see, e.g., \citet{Terquem:21, Terquem:21b}, but also \cite{Barker:21}). Another possibility is that the planet has highly dissipative g-modes either in the surface layers~\citep{Jermyn:17} or in the core~\citep{Mankovich:21}.  Alternatively, modes in the host star might become important through either resonant locking~\citep{Ma:21} or the parameteric instability~\citep{Essick:16}.

%One immediate prediction of our calculation is that there should be a piling-up of systems at the termination points of the diffusive tide. This is because the orbital decay due to the diffusive tide is highly efficient (happening on a timescale of $\sim 10\,{\rm kyr}$) compared to the subsequent evolution requiring 10 to 100 Myr (see, e.g., \citealt{Vick:19}, though their calculation was based on a parameterized constant tidal lag). This suggests that under the linear theory alone, the fact that many systems should stall at $e>0.9$ (Figures~\ref{fig:terminal_pt_strongly_damped} and \ref{fig:terminal_pt_weakly_damped}) would render it in tension with the observed paucity of super-eccentric ($e> 0.9$) proto-hot Jupiters \citep{Socrates:12, Dawson:15}. Including nonlinear mode interactions, on the other hand, mitigates the tension slightly by allowing the orbit to circularize further to $e\simeq 0.8$. The current observational constraints on proto-hot Jupiters with $0.6<e<0.9$ is not well established because such systems are hard to be identified using the photoeccentric effect~\citep{Dawson:15}. Our study thus serves as a motivation for novel searching techniques and strategies for proto-hot Jupiters in the $0.6<e<0.9$ band. 

A limitation of our current model is that it does not account for the evolution of the planet's background structure. In order for the orbit to shrink from $a= 1\,{\rm AU}$ to $0.1\,{\rm AU}$, about $2 E_0$ of orbital energy needs to be dissipated. Depositing this amount of energy in the planet could significantly alter its  structure (HJs with inflated radii are not necessarily explained by such energy deposition since the planet cools after the diffusive tide terminates; see, e.g., \citealt{Baraffe:10} and references therein). These structural changes may in turn feed back on the diffusive tidal evolution. Meanwhile, if the energy is all dissipated as heat, the mean luminosity of the planet could reach $\sim 5\% L_\odot \sim 10^{-3} L_{\rm Edd}$, where $L_{\rm Edd}$ is the Eddington luminosity of the planet. This is comparable to the luminosity of a $0.5\,M_\odot$ main-sequence star. If the energy dissipates through sudden wave-breaking events, the instantaneous luminosity could even be  super-Eddington for brief periods  and launch radiation-driven winds  \citep{Wu:18}.

Another limitation of our analysis  is that for simplicity we  assume the  planet's rotation rate is constant throughout the diffusive tidal  evolution.  In effect, we treat the f-mode as a ``sink'' of orbital AM and do not consider the subsequent transfer of the AM from the f-mode to the background planet, nor the further redistribution of the AM in the planet. Mode-spin coupling and AM transport  are challenging problems. In Appendix~\ref{appx:spin_effects}, we consider a limiting case where the AM transferred from the orbit to the mode is instantaneously deposited to the background planet and the planet, in turn, instantaneously transports the AM such that it rotates as a solid body throughout the evolution. In this limit, we find that the planet could be spun up significantly, with $\Omega_s > \omega_0$ if it absorbs all the orbital energy and associated orbital AM from $a= 1\,{\rm AU}$ to $0.1\,{\rm AU}$. In reality, before the planet could ever reach such a high spin rate, the diffusive evolution of the f-mode would likely be quenched.  This is because the inertial-frame mode frequency $\omega_a'$ then becomes so large that the tidal kick amplitude diminishes  significantly  (Equation~(\ref{eq:K_lm_para})). Such a scenario might be avoided if there is a significant mass outflow that carries away the excess AM.  The orbital decay timescale under the diffusive tide would then likely be limited by the spin-down timescale.   Alternatively, planetary inertial modes could be excited if the planet rotates near the critical spin rate~\citep{Papaloizou:05, Ivanov:07, Xu:17, Vick:19}, which may allow the diffusive tidal evolution to persist. Understanding the spin evolution of the planet during diffusive growth is an important topic that future studies will need to address.

% \begin{acknowledgements}
% We thank Fei Dai for helpful discussions and comments during the preparation of the manuscript.
% This work was supported by NSF AST-2054353.
% H.Y. acknowledges the support of the Sherman Fairchild Foundation.  
% \end{acknowledgements}

%%%% for arXiv version, no internal line numbers %%%
We thank Fei Dai for helpful discussions and comments during the preparation of the manuscript.
This work was supported by NSF AST-2054353.
H.Y. acknowledges the support of the Sherman Fairchild Foundation.

%\clearpage
\appendix

\section{Diffusive growth threshold as a function of final orbital period and planetary radius}
\label{appx:P_f_vs_R}

In the analysis presented in Section~\ref{sec:number_along_track}, we grouped systems only according to their final orbital period $P_{\rm orb}^{\rm (f)}$. While such a simple grouping is consistent with the analysis performed by \citet{Socrates:12} and is typically done observationally (e.g., \citealt{Dawson:15}), we point out here as a caveat that it may oversimplify the problem. 

For example, the number density can also be a sensitive function of the planetary radius $R$. In part, this is because $R$ would enter the overall timescale $T(e)$ as $R^{-5}$ (Equation~\ref{eq:T_tidal_scale}). More importantly, the tidal lag time $\tau$ in Equation~(\ref{eq:tau_tidal_lag}) should be treated as a function of $R$ as well, because the threshold for diffusive growth to happen is sensitive to $R$. 

\begin{figure}
   \centering
   \includegraphics[width=0.45\textwidth]{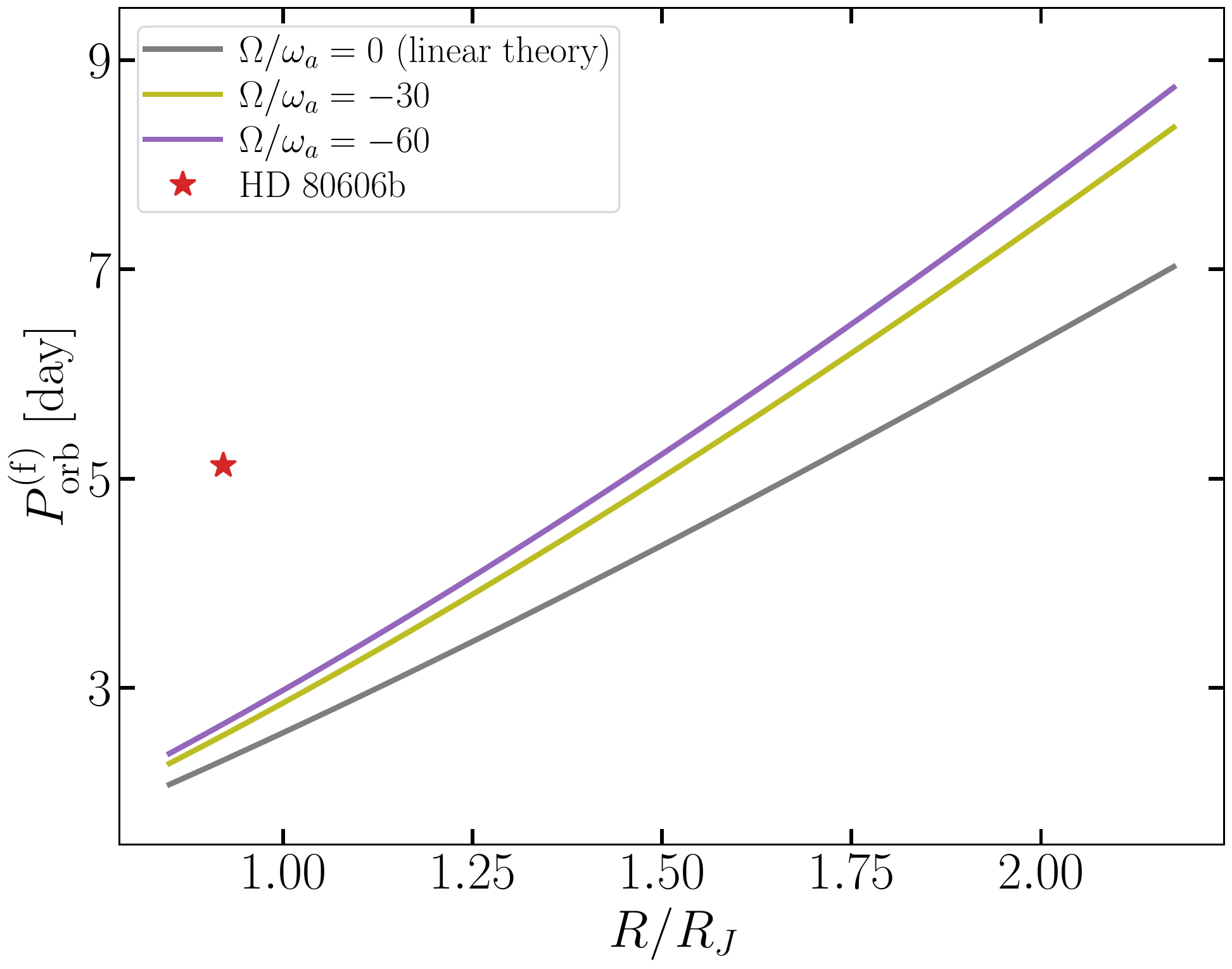} % requires the graphicx package
\caption{Maximum $P_{\rm orb}^{\rm (f)}$ at which the diffusive tide can still be triggered as a function of the planetary radius. The gray line corresponds to the threshold calculated under the linear theory, while the olive and purple lines include the phase shift induced by nonlinear mode interactions assuming different values of $\Omega/\omega_a$ (Table~\ref{tab:models}). The system HD 80606b (red star) is above the threshold and therefore will not undergo diffusive growth.}
\label{fig:P_f_vs_R}
\end{figure}

We explore this point in detail in Figure~\ref{fig:P_f_vs_R}. Here we show the maximum $P_{\rm orb}^{\rm (f)}$ at which  diffusive growth can happen as a function of $R$. Our calculation follows section 5.1 in \citet{Yu:21}. To generate the curves, we fix $(M, M_\ast)=(M_J, M_\odot)$ and $a=1\,{\rm AU}$. We further set $\omega_a=1.1\omega_0$ and $Q_a=0.4$. The planet is assumed to be non-rotating; including rotation would reduce the maximum $P_{\rm orb}^{\rm (f)}$ allowed further as it reduces the tidal kick at the pericenter. 

% If the diffusive tide is a necessary component for the high-eccentricity tidal migration (as it simultaneously makes the overall circularization timescale shorter than the age of the host star and it pervents the tidal disruption )

One might argue that the diffusive tide is a necessary component for the high-eccentricity tidal migration of proto-HJs. This is because it simultaneously makes the overall circularization timescale shorter than the age of the host star and it prevents the eccentricity from getting excited to too high a value, thereby saving the planet from being disrupted by the host star (see, e.g., the discussion in \citealt{Vick:19}). If this is the case, then Figure~\ref{fig:P_f_vs_R} suggests that the final orbital period of an HJ would be highly correlated with its initial radius, as only Jupiters with $R\gtrsim 1.5 R_J$ could successfully migrate to an orbit with $P_{\rm orb}^{\rm (f)}=5\,{\rm day}$. However, this correlation may be washed out for HJs today because after undergoing diffusive evolution, the planet has sufficient time to cool and contract (and possibly re-inflate), making their current radii differ from their initial ones.

Figure~\ref{fig:P_f_vs_R} further suggests that to make accurate predictions of the eccentricity distribution, more careful studies that include population synthesis and planetary evolution are necessary.

\section{Tidal angular momentum transfer}
\label{appx:AM_transfer}
In this appendix, we calculate the amount of AM transferred from the orbit to the modes as a function of the energy transferred to the modes.
The energy deposited from the orbit to the planet is 
\begin{eqnarray}
    &\dot{E}(t) &= -\int d^3 \vect{x} \rho \dot{\vect{\xi}} \cdot \nabla U \nonumber \\
    &&=\sum_a (i\omega_a) q_a^\ast(t) U_a(t) E_0, \nonumber \\
    &&=-2\sum_{\omega_a>0} {\rm Im}\left[\omega_a q_a^\ast U_a\right] E_0,
\end{eqnarray}
where in the second line we have used the modal expansion to expend the velocity $\dot{\vect{\xi}}$ into eigenmodes.

The torque $\mathcal{T}$ acting on the planet can be derived from an interaction Hamiltonian (see, e.g., \citealt{Yu:20a}), 
\begin{equation}
    H_{\rm int} = \sum_{\omega_a>0} \left[q_a U_a^\ast + q_a^\ast U_a)\right]E_0,
\end{equation}
leading to
\begin{eqnarray}
    &\mathcal{T} & = \frac{\partial H_{\rm int}}{\partial \Phi} =  -2\sum_{\omega_a > 0} {\rm Re}\left[q_a^\ast \frac{\partial U_a}{\partial \Phi}\right] E_0,\nonumber \\
    && = -2 \sum_{\omega_a > 0}{\rm Im}\left[m_a q_a^\ast U_a\right] E_0. 
\end{eqnarray}

Thus the energy transfer to a mode $\Delta E_a$ is related to the AM transfer to the mode $\Delta J_a$  by 
\begin{equation}
    \Delta E_a = \frac{\omega_a}{m_a} \Delta J_a.
    \label{eq:dE_vs_dJ}
\end{equation}
The same relation can also be derived from the Lagrangian of the perturbed fluid. See, e.g., \citet{Friedman:78} for details. 

It is also interesting to consider how the energy/AM transfer affects the orbit (see Equation~(\ref{eq:dJ_eq_dL})). This can be done by dividing both sides of Equation~(\ref{eq:dE_vs_dJ}) by $|E_{\rm orb}| = \Omega_{\rm orb} L / 2$, leading to 
\begin{equation}
    \frac{\Delta E_a}{|E_{\rm orb}|} = \frac{2\omega_a}{m_a\Omega_{\rm orb}} \frac{\Delta J_a}{L}. 
\end{equation}
Therefore, for a highly eccentric orbit with $\omega_a \left(\gtrsim \Omega_{\rm peri}\right)\gg\Omega_{\rm orb}$, the fractional change in the orbital AM $L$ is much smaller than the fractional change in the orbital energy $E_{\rm orb}$.

\section{Properties of the $1.7R_J$ Jupiter model}
\label{appx:1p7Rj_Jupiter}

We show here the properties of the daughter modes in the $1.7R_J$ model. The calculations closely follow those of section 3.3 in \citet{Yu:21}. 

\begin{figure}
   \centering
   \includegraphics[width=0.45\textwidth]{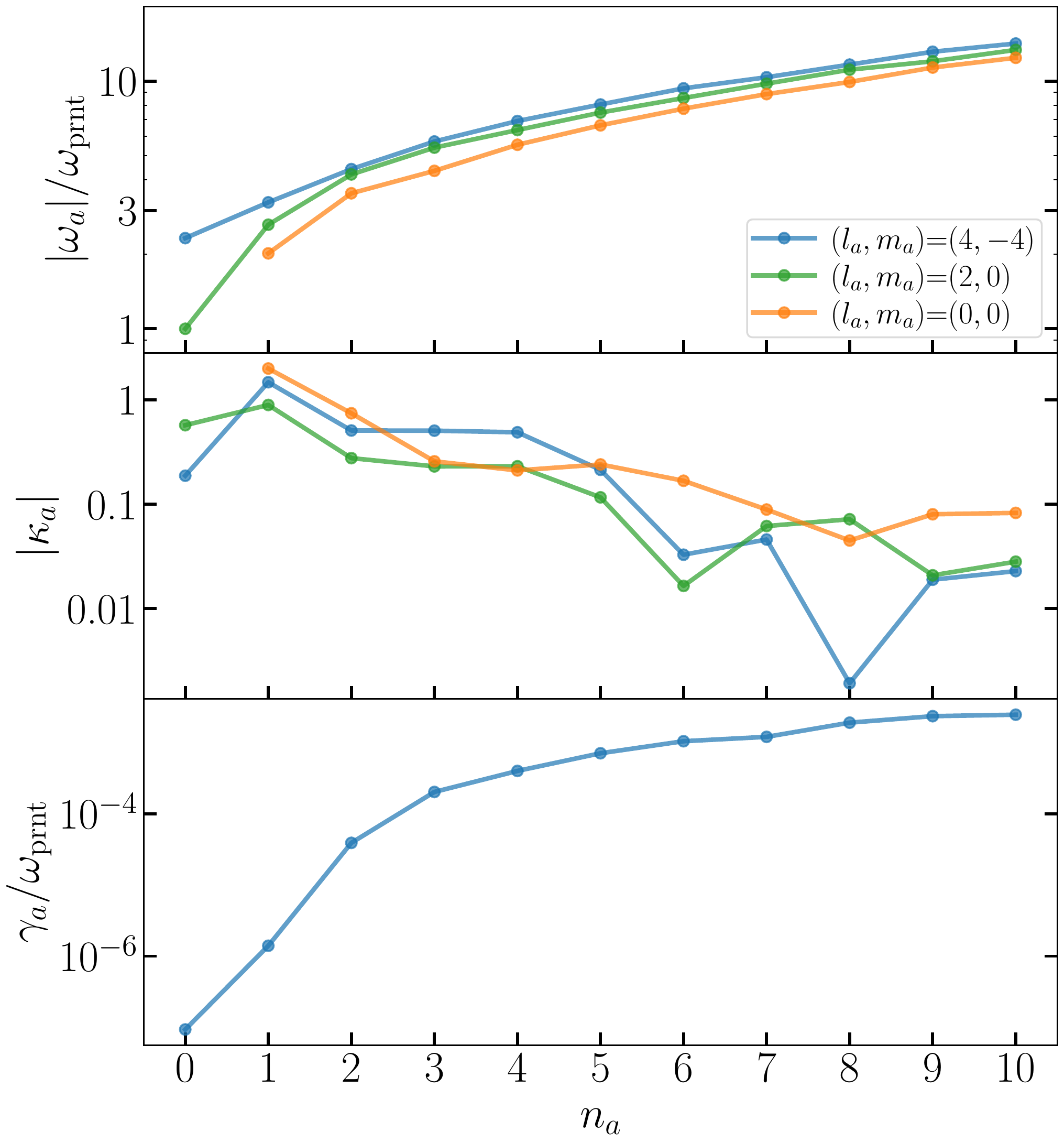} % requires the graphicx package
\caption{Eigenfrequency, coupling coefficient, and linear damping rate of modes of the $R=1.7R_J$ model as a function of their radial order $n_a$ (see key in the top panel for their $l_a$ and $m_a$). Here we use the subscript ``$a$'' to stand for a generic mode, and we use $\omega_{\rm prnt}$ to indicate specifically the linear eigenfrequency of the parent (i.e., the frequency given in Table~\ref{tab:models}).  %\phil{($\omega_{\rm prnt}$ is the linear eigenfrequency of the parent. Say this in caption rather than text.)} 
}
\label{fig:lin_omega_gam_1p0Mj_1p7Rj}
\end{figure}

In Figure~\ref{fig:lin_omega_gam_1p0Mj_1p7Rj}, we show the linear frequency, the coupling coefficient with the parent mode \citep{Weinberg:12}, and the linear damping rate due to turbulent convection of each daughter mode~\citep{Shiode:12, Burkart:13}.

\begin{figure}
   \centering
   \includegraphics[width=0.45\textwidth]{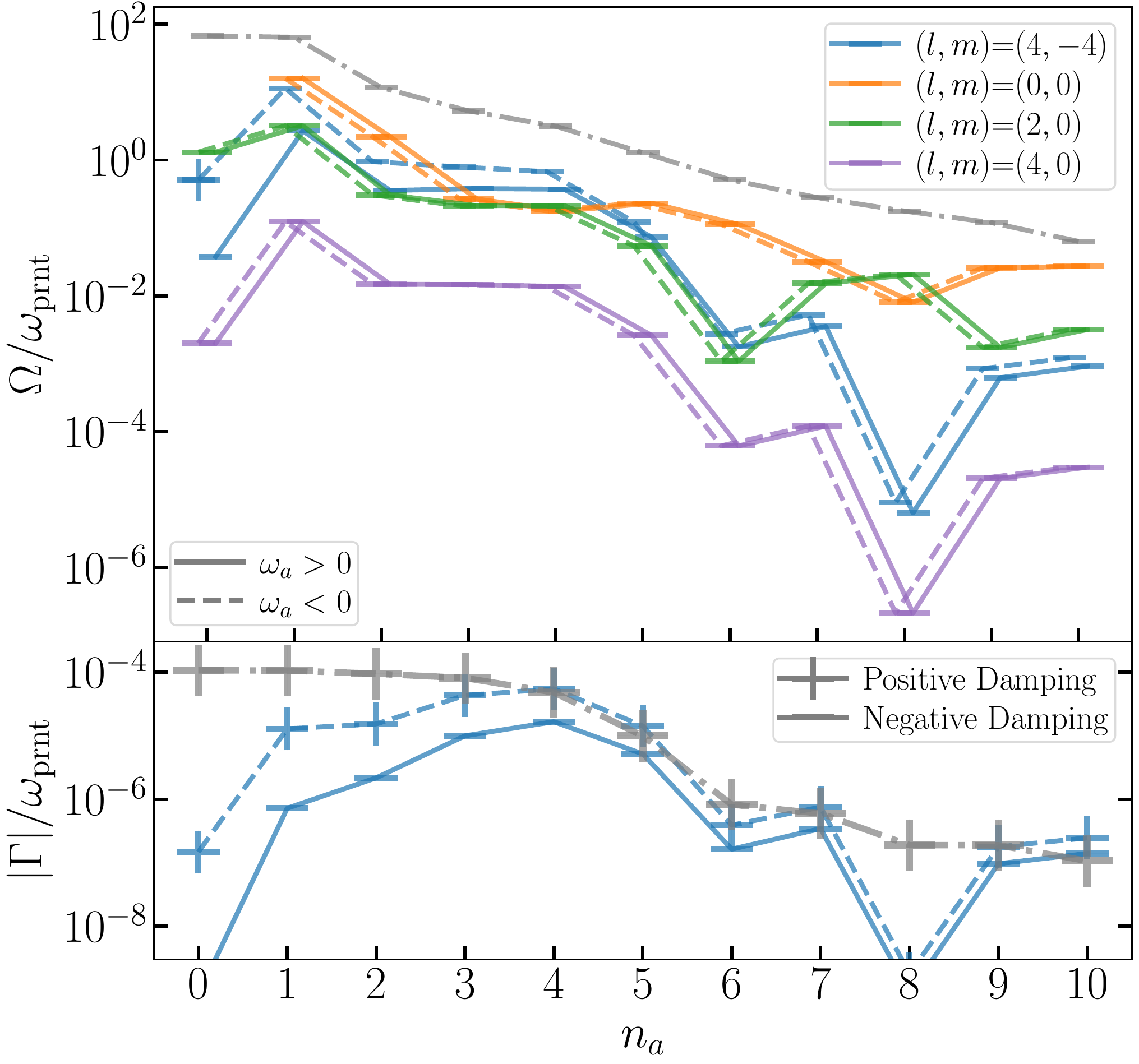} % requires the graphicx package
\caption{Each linear eigenmode's contribution to the nonlinear frequency shift parameter $\Omega$ (top) and dissipation rate parameter $\Gamma$ (bottom). }
\label{fig:Omega_Gamma_1p0Mj_1p7Rj}
\end{figure}

From the linear properties of the daughter modes, we can compute their contributions to the nonlinear frequency shift $\Omega$ and damping rate $\Gamma$ following \citet{Yu:21}. The result is shown in Figure~\ref{fig:Omega_Gamma_1p0Mj_1p7Rj}. Here in the colored curves, we show each mode's contribution, with a ``$+(-)$'' symbol indicating a positive (negative) value. In the gray curves, we plot the cumulative contributions to $\Omega$ and $\Gamma$ from modes with radial order $\geq n_a$ and all values of $(l_a, m_a)$ allowed by the angular selection rules~\citep{Weinberg:12}. 

\section{Spin as an alternative channel for terminating diffusive growth}
\label{appx:spin_effects}

\begin{figure}
   \centering
   \includegraphics[width=0.45\textwidth]{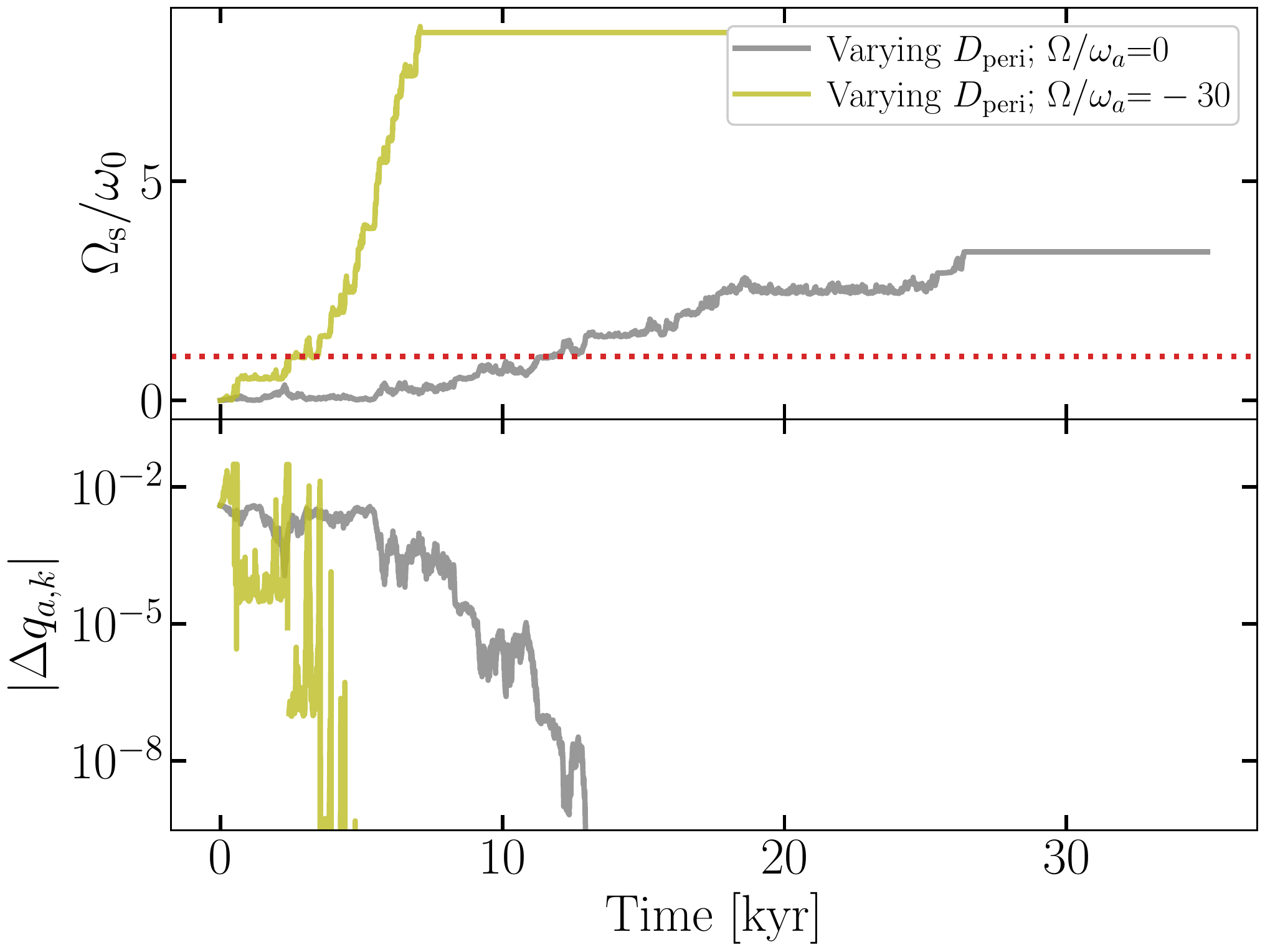} % requires the graphicx package
\caption{Spin evolution (top panel) and the kick amplitude (bottom panel) found by post-processing the trajectories shown in Figure~\ref{fig:kyr_evol_traj_R_1p1Rj}. %\nevin{Maybe end caption here since already say the interpretation that follows in the text.} Here we consider a limiting case where all the orbital angular momentum transferred to the f-mode gets instantaneously deposited into the background planet, and that the planet always maintains solid-body rotation. Because the total orbital energy that needs to be dissipated exceeds the binding energy of the planet $E_0$, the spin of the planet can also exceed $\omega_0$ if the angular momentum is not somehow lost. In fact, in this efficient AM transporting limit, the diffusion will be terminated shortly after its onset, which thus prevents the spin from reaching unphysically large values.  This is illustrated in the bottom panel where we show the magnitude of the kick including the Doppler shift of the mode frequency in the inertial frame.
}
\label{fig:kyr_evol_omega_s_R_1p1Rj}
\end{figure}

In the main text, we assumed that the f-mode acts as the ultimate sink of the AM. Here we consider the possibility that the orbital AM transferred into the f-mode then gets deposited in the background planet and spins it up.  For simplicity, we assume that the transfer from the mode to the planet is instantaneous and that the planet maintains a solid-body rotation at all times. Under these assumptions, we show that the spin-up of the background planet can be another process by which the diffusive growth terminates.

In Figure~\ref{fig:kyr_evol_omega_s_R_1p1Rj} we show the spin evolution $\Omega_s(t)$  found by \emph{post-processing} the trajectories in Figure~\ref{fig:kyr_evol_traj_R_1p1Rj}. In other words, we use the $\Delta J_a$ computed when generating Figure~\ref{fig:kyr_evol_traj_R_1p1Rj} %\nevin{do you mean when generating Figure~\ref{fig:kyr_evol_traj_R_1p1Rj}?} 
to calculate the spin rate of the planet at the $k$'th orbit according to $\Omega_{s, k}=\sum_i^k \Delta J_{a, i}/I$, where $I=0.38 MR^2$ is the moment of inertia of the $1.1 R_J$ model. Once we have the spin of the background planet, we can also calculate the kick amplitude at each pericenter passage according to Equation~(\ref{eq:K_lm_para}) with the new $\omega_{a, k}'=\omega_a + m_a \Omega_{s,k}$. The result is shown in the lower panel in Figure~\ref{fig:kyr_evol_omega_s_R_1p1Rj}. Note, however, that since this is done in post-processing we do not further change the evolution trajectories based on the new $\Delta q_{a,k}$. 

The top panel shows that the planet spins-up significantly.  This is because in order for the orbit to shrink from $a\simeq 1\,{\rm AU}$ to $\simeq 0.1\,{\rm AU}$, the total amount of energy that needs to be transferred into the planet and dissipated can be  few times $E_0$. From Equation~(\ref{eq:dE_vs_dJ}), the total amount of orbital AM absorbed is 
\begin{equation}
    \frac{\Delta J_a}{J_0} = \left(\frac{m_a\omega_0}{\omega_a}\right)\left(\frac{\Delta E_a}{E_0}\right),
\end{equation}
where $J_0\equiv \omega_0 MR^2$. 

In fact, before the planet reaches such a large spin rate, the diffusive evolution is likely to terminate because of the sharp decrease of $|\Delta q_{a, k}|$ with increasing $\omega_a'$, as indicated in the bottom panel (see also Equation~(\ref{eq:K_lm_para})). If the AM gained by the background planet cannot be removed somehow, this would limit the total amount of orbital energy decay to be $\ll E_0$. 
%Further orbital evolution via the diffusive tide would thus invoke to remove the excess AM and slows down the spin of the planet.
One potential way to prevent such a large spin-up and thereby delay the termination of the diffusive evolution is if the planet drives a wind that induces a strong  spin-down torque
(e.g. a magnetic Weber-Davis wind;~\citealt{Weber:67}). 
For moment-arm $R_{\rm A}$, the AM loss rate scales as 
\begin{equation}
    \dot{J} = \frac{2}{3}\dot{M}\Omega_s R_{\rm A}^2,
\end{equation}
where $\dot{M}$ is the mass loss rate. 
We can thus define a spin-down timescale $\tau_{\rm sd}$ as the time for the wind to remove $J_0$ from the planetary AM,  
\begin{eqnarray}
    &&\tau_{\rm sd}=\frac{J_0}{\dot{J}}=\frac{3}{2}
    \left(\frac{M}{\dot{M}}\right)
    \left(\frac{R}{R_{\rm A}}\right)^2
    \left(\frac{\omega_0}{\Omega_s}\right),
    \nonumber \\
    &&\simeq 2\times10^4 {\rm yr}
    \left(\frac{\dot{M}/M}{10^{-5}\,{\rm yr}^{-1}}\right)^{-1}
    \left(\frac{R_{\rm A}/R}{3}\right)^{-2}
    \left(\frac{\omega_0}{\Omega_s}\right),
\end{eqnarray}
where the co-rotating moment arm (determined by, e.g., the magnetosphere) has been assumed to extend to $R_{\rm A}=3R$. 
In other words, the planet needs to have a significant mass loss rate of $\dot{M}/M\sim 10^{-5}\,{\rm yr^{-1}}$ in order to keep the spin rate low enough that the diffusive evolution can continue unabated.

In the absence of such a powerful outflow, the circularization timescale will be limited by the spin-down timescale, if in fact there is any signficiant spin-down mechanism. In this case, the diffusive tide could only operate momentarily before it is shut down by the rapid spin of the planet. We would then need to wait for the mass outflow to slowly remove the AM until $|\Delta q_{a,k}|$ becomes large to be able to trigger diffusion again. The time to circularize the orbit can then be no faster than the spin-down timescale $\tau_{\rm sd}$. . 

%In either case,\nevin{what are the two cases? with and without a powerful outflow?} 
Regardless of the mass loss rate,
the total amount of mass that needs to be removed is given by 
\begin{eqnarray}
    &&\frac{\Delta M}{M}=\frac{3}{2}
    \left(\frac{\Delta J_a}{J_0}\right)
    \left(\frac{R}{R_{\rm A}}\right)^2
    \left(\frac{\omega_0}{\Omega_s}\right). 
\end{eqnarray}
Therefore, a significant amount of mass and angular momentum  loss is likely required in order for the f-mode to experience a significant diffusive growth. 
%We conclude, therefore, that there must be a significant amount of mass and angular momentum loss if  diffusive growth is to occur.\nevin{Is this too strong a conclusion given the assumptions we made? For example, if we don't assume solid body rotation (although the planet is convective) or if we account for inertial modes, could the demand on AM and mass loss to maintain diffusive growth be somewhat less severe?}  
%and it could potentially affect, e.g., the composition of the planet's atmosphere. 

%where we have assumed a mass loss rate of $\dot{M}\simeq 10^{-13} M_J\,{\rm yr}^{-1}\simeq 6\times10^{9} {\rm g\,s^{-1}}$~\citep{Cherenkov:14}. From this estimation we see that unless the mass loss rate is much more efficient than the one assumed,  and/or the Alfv\'en radius $R_{\rm A}\gg30R_J$, the wind is likely to be too inefficient to remove the AM. 


\bibliography{ref}{}

\begin{thebibliography}{}
\expandafter\ifx\csname natexlab\endcsname\relax\def\natexlab#1{#1}\fi
\providecommand{\url}[1]{\href{#1}{#1}}
\providecommand{\dodoi}[1]{doi:~\href{http://doi.org/#1}{\nolinkurl{#1}}}
\providecommand{\doeprint}[1]{\href{http://ascl.net/#1}{\nolinkurl{http://ascl.net/#1}}}
\providecommand{\doarXiv}[1]{\href{https://arxiv.org/abs/#1}{\nolinkurl{https://arxiv.org/abs/#1}}}

\bibitem[{{Baraffe} {et~al.}(2010){Baraffe}, {Chabrier}, \&
  {Barman}}]{Baraffe:10}
{Baraffe}, I., {Chabrier}, G., \& {Barman}, T. 2010, Reports on Progress in
  Physics, 73, 016901, \dodoi{10.1088/0034-4885/73/1/016901}

\bibitem[{{Barker} \& {Astoul}(2021)}]{Barker:21}
{Barker}, A.~J., \& {Astoul}, A. A.~V. 2021, \mnras, 506, L69,
  \dodoi{10.1093/mnrasl/slab077}

\bibitem[{{Burkart} {et~al.}(2013){Burkart}, {Quataert}, {Arras}, \&
  {Weinberg}}]{Burkart:13}
{Burkart}, J., {Quataert}, E., {Arras}, P., \& {Weinberg}, N.~N. 2013, \mnras,
  433, 332, \dodoi{10.1093/mnras/stt726}

\bibitem[{{Burke} {et~al.}(2014){Burke}, {Bryson}, {Mullally}, {Rowe},
  {Christiansen}, {Thompson}, {Coughlin}, {Haas}, {Batalha}, {Caldwell},
  {Jenkins}, {Still}, {Barclay}, {Borucki}, {Chaplin}, {Ciardi}, {Clarke},
  {Cochran}, {Demory}, {Esquerdo}, {Gautier}, {Gilliland}, {Girouard}, {Havel},
  {Henze}, {Howell}, {Huber}, {Latham}, {Li}, {Morehead}, {Morton}, {Pepper},
  {Quintana}, {Ragozzine}, {Seader}, {Shah}, {Shporer}, {Tenenbaum}, {Twicken},
  \& {Wolfgang}}]{Burke:14}
{Burke}, C.~J., {Bryson}, S.~T., {Mullally}, F., {et~al.} 2014, \apjs, 210, 19,
  \dodoi{10.1088/0067-0049/210/2/19}

\bibitem[{{Chatterjee} {et~al.}(2008){Chatterjee}, {Ford}, {Matsumura}, \&
  {Rasio}}]{Chatterjee:08}
{Chatterjee}, S., {Ford}, E.~B., {Matsumura}, S., \& {Rasio}, F.~A. 2008, \apj,
  686, 580, \dodoi{10.1086/590227}

\bibitem[{{Darwin}(1879)}]{Darwin:1879}
{Darwin}, G.~H. 1879, Philosophical Transactions of the Royal Society of London
  Series I, 170, 1

\bibitem[{{Dawson} {et~al.}(2015){Dawson}, {Murray-Clay}, \&
  {Johnson}}]{Dawson:15}
{Dawson}, R.~I., {Murray-Clay}, R.~A., \& {Johnson}, J.~A. 2015, \apj, 798, 66,
  \dodoi{10.1088/0004-637X/798/2/66}

\bibitem[{{Dong} {et~al.}(2021){Dong}, {Huang}, {Zhou}, {Dawson}, {Rodriguez},
  {Eastman}, {Collins}, {Quinn}, {Shporer}, {Triaud}, {Wang}, {Beatty},
  {Jackson}, {Collins}, {Abe}, {Suarez}, {Crouzet}, {M{\'e}karnia},
  {Dransfield}, {Jensen}, {Stockdale}, {Barkaoui}, {Heitzmann}, {Wright},
  {Addison}, {Wittenmyer}, {Okumura}, {Bowler}, {Horner}, {Kane}, {Kielkopf},
  {Liu}, {Plavchan}, {Mengel}, {Ricker}, {Vanderspek}, {Latham}, {Seager},
  {Winn}, {Jenkins}, {Christiansen}, \& {Paegert}}]{Dong:21}
{Dong}, J., {Huang}, C.~X., {Zhou}, G., {et~al.} 2021, \apjl, 920, L16,
  \dodoi{10.3847/2041-8213/ac2600}

\bibitem[{{Essick} \& {Weinberg}(2016)}]{Essick:16}
{Essick}, R., \& {Weinberg}, N.~N. 2016, \apj, 816, 18,
  \dodoi{10.3847/0004-637X/816/1/18}

\bibitem[{{Fabrycky} \& {Tremaine}(2007)}]{Fabrycky:07}
{Fabrycky}, D., \& {Tremaine}, S. 2007, \apj, 669, 1298, \dodoi{10.1086/521702}

\bibitem[{{Friedman} \& {Schutz}(1978)}]{Friedman:78}
{Friedman}, J.~L., \& {Schutz}, B.~F. 1978, \apj, 221, 937,
  \dodoi{10.1086/156098}

\bibitem[{{Goldreich} {et~al.}(1989){Goldreich}, {Murray}, {Longaretti}, \&
  {Banfield}}]{Goldreich:89}
{Goldreich}, P., {Murray}, N., {Longaretti}, P.~Y., \& {Banfield}, D. 1989,
  Science, 245, 500, \dodoi{10.1126/science.245.4917.500}

\bibitem[{{Goldreich} \& {Nicholson}(1977)}]{Goldreich:77}
{Goldreich}, P., \& {Nicholson}, P.~D. 1977, \icarus, 30, 301,
  \dodoi{10.1016/0019-1035(77)90163-4}

\bibitem[{{Goldreich} \& {Soter}(1966)}]{Goldreich:66}
{Goldreich}, P., \& {Soter}, S. 1966, \icarus, 5, 375,
  \dodoi{10.1016/0019-1035(66)90051-0}

\bibitem[{{Hamers} {et~al.}(2017){Hamers}, {Antonini}, {Lithwick}, {Perets}, \&
  {Portegies Zwart}}]{Hamers:17}
{Hamers}, A.~S., {Antonini}, F., {Lithwick}, Y., {Perets}, H.~B., \& {Portegies
  Zwart}, S.~F. 2017, \mnras, 464, 688, \dodoi{10.1093/mnras/stw2370}

\bibitem[{{Hut}(1981)}]{Hut:81}
{Hut}, P. 1981, \aap, 99, 126

\bibitem[{{Ivanov} \& {Papaloizou}(2004)}]{Ivanov:04}
{Ivanov}, P.~B., \& {Papaloizou}, J.~C.~B. 2004, \mnras, 347, 437,
  \dodoi{10.1111/j.1365-2966.2004.07238.x}

\bibitem[{{Ivanov} \& {Papaloizou}(2007)}]{Ivanov:07}
---. 2007, \mnras, 376, 682, \dodoi{10.1111/j.1365-2966.2007.11463.x}

\bibitem[{{Jermyn} {et~al.}(2017){Jermyn}, {Tout}, \& {Ogilvie}}]{Jermyn:17}
{Jermyn}, A.~S., {Tout}, C.~A., \& {Ogilvie}, G.~I. 2017, \mnras, 469, 1768,
  \dodoi{10.1093/mnras/stx831}

\bibitem[{{Lai}(1997)}]{Lai:97}
{Lai}, D. 1997, \apj, 490, 847, \dodoi{10.1086/304899}

\bibitem[{{Ma} \& {Fuller}(2021)}]{Ma:21}
{Ma}, L., \& {Fuller}, J. 2021, arXiv e-prints, arXiv:2105.09335.
\newblock \doarXiv{2105.09335}

\bibitem[{{Mankovich} \& {Fuller}(2021)}]{Mankovich:21}
{Mankovich}, C., \& {Fuller}, J. 2021, arXiv e-prints, arXiv:2104.13385.
\newblock \doarXiv{2104.13385}

\bibitem[{{Mardling}(1995)}]{Mardling:95}
{Mardling}, R.~A. 1995, \apj, 450, 722, \dodoi{10.1086/176178}

\bibitem[{{Mu{\~n}oz} {et~al.}(2016){Mu{\~n}oz}, {Lai}, \& {Liu}}]{Munoz:16}
{Mu{\~n}oz}, D.~J., {Lai}, D., \& {Liu}, B. 2016, \mnras, 460, 1086,
  \dodoi{10.1093/mnras/stw983}

\bibitem[{{Ogilvie}(2014)}]{Ogilvie:14}
{Ogilvie}, G.~I. 2014, \araa, 52, 171,
  \dodoi{10.1146/annurev-astro-081913-035941}

\bibitem[{{Ogilvie} \& {Lin}(2004)}]{Ogilvie:04}
{Ogilvie}, G.~I., \& {Lin}, D.~N.~C. 2004, \apj, 610, 477,
  \dodoi{10.1086/421454}

\bibitem[{{Papaloizou} \& {Ivanov}(2005)}]{Papaloizou:05}
{Papaloizou}, J.~C.~B., \& {Ivanov}, P.~B. 2005, \mnras, 364, L66,
  \dodoi{10.1111/j.1745-3933.2005.00107.x}

\bibitem[{{Paxton} {et~al.}(2011){Paxton}, {Bildsten}, {Dotter}, {Herwig},
  {Lesaffre}, \& {Timmes}}]{Paxton:11}
{Paxton}, B., {Bildsten}, L., {Dotter}, A., {et~al.} 2011, \apjs, 192, 3,
  \dodoi{10.1088/0067-0049/192/1/3}

\bibitem[{{Paxton} {et~al.}(2013){Paxton}, {Cantiello}, {Arras}, {Bildsten},
  {Brown}, {Dotter}, {Mankovich}, {Montgomery}, {Stello}, {Timmes}, \&
  {Townsend}}]{Paxton:13}
{Paxton}, B., {Cantiello}, M., {Arras}, P., {et~al.} 2013, \apjs, 208, 4,
  \dodoi{10.1088/0067-0049/208/1/4}

\bibitem[{{Paxton} {et~al.}(2015){Paxton}, {Marchant}, {Schwab}, {Bauer},
  {Bildsten}, {Cantiello}, {Dessart}, {Farmer}, {Hu}, {Langer}, {Townsend},
  {Townsley}, \& {Timmes}}]{Paxton:15}
{Paxton}, B., {Marchant}, P., {Schwab}, J., {et~al.} 2015, \apjs, 220, 15,
  \dodoi{10.1088/0067-0049/220/1/15}

\bibitem[{{Paxton} {et~al.}(2018){Paxton}, {Schwab}, {Bauer}, {Bildsten},
  {Blinnikov}, {Duffell}, {Farmer}, {Goldberg}, {Marchant}, {Sorokina},
  {Thoul}, {Townsend}, \& {Timmes}}]{Paxton:18}
{Paxton}, B., {Schwab}, J., {Bauer}, E.~B., {et~al.} 2018, \apjs, 234, 34,
  \dodoi{10.3847/1538-4365/aaa5a8}

\bibitem[{{Paxton} {et~al.}(2019){Paxton}, {Smolec}, {Schwab}, {Gautschy},
  {Bildsten}, {Cantiello}, {Dotter}, {Farmer}, {Goldberg}, {Jermyn}, {Kanbur},
  {Marchant}, {Thoul}, {Townsend}, {Wolf}, {Zhang}, \& {Timmes}}]{Paxton:19}
{Paxton}, B., {Smolec}, R., {Schwab}, J., {et~al.} 2019, \apjs, 243, 10,
  \dodoi{10.3847/1538-4365/ab2241}

\bibitem[{{Petrovich} \& {Tremaine}(2016)}]{Petrovich:16}
{Petrovich}, C., \& {Tremaine}, S. 2016, \apj, 829, 132,
  \dodoi{10.3847/0004-637X/829/2/132}

\bibitem[{{Press} \& {Teukolsky}(1977)}]{Press:77}
{Press}, W.~H., \& {Teukolsky}, S.~A. 1977, \apj, 213, 183,
  \dodoi{10.1086/155143}

\bibitem[{{Rasio} \& {Ford}(1996)}]{Rasio:96}
{Rasio}, F.~A., \& {Ford}, E.~B. 1996, Science, 274, 954,
  \dodoi{10.1126/science.274.5289.954}

\bibitem[{{Schenk} {et~al.}(2002){Schenk}, {Arras}, {Flanagan}, {Teukolsky}, \&
  {Wasserman}}]{Schenk:02}
{Schenk}, A.~K., {Arras}, P., {Flanagan}, {\'E}.~{\'E}., {Teukolsky}, S.~A., \&
  {Wasserman}, I. 2002, \prd, 65, 024001, \dodoi{10.1103/PhysRevD.65.024001}

\bibitem[{{Shiode} {et~al.}(2012){Shiode}, {Quataert}, \& {Arras}}]{Shiode:12}
{Shiode}, J.~H., {Quataert}, E., \& {Arras}, P. 2012, \mnras, 423, 3397,
  \dodoi{10.1111/j.1365-2966.2012.21130.x}

\bibitem[{{Socrates} {et~al.}(2012){Socrates}, {Katz}, {Dong}, \&
  {Tremaine}}]{Socrates:12}
{Socrates}, A., {Katz}, B., {Dong}, S., \& {Tremaine}, S. 2012, \apj, 750, 106,
  \dodoi{10.1088/0004-637X/750/2/106}

\bibitem[{{Terquem}(2021{\natexlab{a}})}]{Terquem:21}
{Terquem}, C. 2021{\natexlab{a}}, \mnras, 503, 5789,
  \dodoi{10.1093/mnras/stab224}

\bibitem[{{Terquem}(2021{\natexlab{b}})}]{Terquem:21b}
---. 2021{\natexlab{b}}, arXiv e-prints, arXiv:2106.15547.
\newblock \doarXiv{2106.15547}

\bibitem[{{Teyssandier} {et~al.}(2013){Teyssandier}, {Naoz}, {Lizarraga}, \&
  {Rasio}}]{Teyssandier:13}
{Teyssandier}, J., {Naoz}, S., {Lizarraga}, I., \& {Rasio}, F.~A. 2013, \apj,
  779, 166, \dodoi{10.1088/0004-637X/779/2/166}

\bibitem[{{Townsend} {et~al.}(2018){Townsend}, {Goldstein}, \&
  {Zweibel}}]{Townsend:18}
{Townsend}, R.~H.~D., {Goldstein}, J., \& {Zweibel}, E.~G. 2018, \mnras, 475,
  879, \dodoi{10.1093/mnras/stx3142}

\bibitem[{{Townsend} \& {Teitler}(2013)}]{Townsend:13}
{Townsend}, R.~H.~D., \& {Teitler}, S.~A. 2013, \mnras, 435, 3406,
  \dodoi{10.1093/mnras/stt1533}

\bibitem[{{Vick} \& {Lai}(2018)}]{Vick:18}
{Vick}, M., \& {Lai}, D. 2018, \mnras, 476, 482, \dodoi{10.1093/mnras/sty225}

\bibitem[{{Vick} {et~al.}(2019){Vick}, {Lai}, \& {Anderson}}]{Vick:19}
{Vick}, M., {Lai}, D., \& {Anderson}, K.~R. 2019, \mnras, 484, 5645,
  \dodoi{10.1093/mnras/stz354}

\bibitem[{{Weber} \& {Davis}(1967)}]{Weber:67}
{Weber}, E.~J., \& {Davis}, Leverett, J. 1967, \apj, 148, 217,
  \dodoi{10.1086/149138}

\bibitem[{{Weinberg} {et~al.}(2012){Weinberg}, {Arras}, {Quataert}, \&
  {Burkart}}]{Weinberg:12}
{Weinberg}, N.~N., {Arras}, P., {Quataert}, E., \& {Burkart}, J. 2012, \apj,
  751, 136, \dodoi{10.1088/0004-637X/751/2/136}

\bibitem[{{Wu}(2018)}]{Wu:18}
{Wu}, Y. 2018, \aj, 155, 118, \dodoi{10.3847/1538-3881/aaa970}

\bibitem[{{Wu} \& {Lithwick}(2011)}]{Wu:11}
{Wu}, Y., \& {Lithwick}, Y. 2011, \apj, 735, 109,
  \dodoi{10.1088/0004-637X/735/2/109}

\bibitem[{{Wu} \& {Murray}(2003)}]{Wu:03}
{Wu}, Y., \& {Murray}, N. 2003, \apj, 589, 605, \dodoi{10.1086/374598}

\bibitem[{{Xu} \& {Lai}(2017)}]{Xu:17}
{Xu}, W., \& {Lai}, D. 2017, \prd, 96, 083005,
  \dodoi{10.1103/PhysRevD.96.083005}

\bibitem[{{Yu} {et~al.}(2021){Yu}, {Weinberg}, \& {Arras}}]{Yu:21}
{Yu}, H., {Weinberg}, N.~N., \& {Arras}, P. 2021, \apj, 917, 31,
  \dodoi{10.3847/1538-4357/ac0a79}

\bibitem[{{Yu} {et~al.}(2020){Yu}, {Weinberg}, \& {Fuller}}]{Yu:20a}
{Yu}, H., {Weinberg}, N.~N., \& {Fuller}, J. 2020, \mnras, 496, 5482,
  \dodoi{10.1093/mnras/staa1858}

\end{thebibliography}
\bibliographystyle{aasjournal}



\end{document}